\documentclass{aa}

\usepackage{graphicx}
\usepackage[outdir=./]{epstopdf}
\usepackage{txfonts}
\usepackage{natbib}
\usepackage{lipsum}
\bibpunct{(}{)}{;}{a}{}{,} 

\begin{document}

\title{Solar Orbiter/RPW antenna calibration in the radio domain and its application to type III burst observations}
        \titlerunning{SO/RPW antenna calibration and type III burst observations}

 \author{A. Vecchio \inst{1,} \inst{2}
        \and  M.Maksimovic \inst{2}
        \and V. Krupar  \inst{3,} \inst{4}
        \and X. Bonnin \inst{2}
        \and A. Zaslavsky\inst{2}
        \and P. L. Astier\inst{2}
        \and M. Dekkali\inst{2}
        \and B. Cecconi\inst{2}
        \and S.D. Bale\inst{5,}\inst{6}
        \and T. Chust\inst{7}
        \and E. Guilhem\inst{8} 
        \and Yu. V. Khotyaintsev\inst{9}
        \and V. Krasnoselskikh\inst{10}
        \and M. Kretzschmar\inst{10}
        \and E. Lorf\`evre\inst{8}
        \and D. Plettemeier\inst{11}
        \and J. Sou\v{c}ek\inst{12}
        \and M. Steller\inst{13} 
        \and \v{S}. \v{S}tver\'ak\inst{14,}\inst{12}  
        \and P. Tr\'avn\'i\v{c}ek\inst{5,}\inst{12}
        \and A. Vaivads\inst{15}
         }
 \institute{
        Radboud Radio Lab, Department of Astrophysics, Radboud University Nijmegen, The Netherlands 
        \and LESIA, Observatoire de Paris, Universit\'e PSL, CNRS, Sorbonne Universit\'e, Universit\'e de Paris, France
        \and Goddard Planetary Heliophysics Institute, University of Maryland,Baltimore County, Baltimore, MD, USA
        \and Heliophysics Science Division, NASA Goddard Space Flight Center, Greenbelt, MD, USA
        \and Space Sciences Laboratory, University of California, Berkeley, CA, USA
        \and Physics Department, University of California, Berkeley, CA, USA
        \and LPP, CNRS, Ecole Polytechnique, Sorbonne Universit\'{e}, Observatoire de Paris, Universit\'{e} Paris-Saclay, Palaiseau, Paris, France         
        \and CNES, 18 Avenue Edouard Belin, 31400 Toulouse, France
        \and Swedish Institute of Space Physics (IRF), Uppsala, Sweden 
        \and  LPC2E, CNRS, 3A avenue de la Recherche Scientifique, Orl\'eans, France  
        \and Technische Universit\"at Dresden, W\"urzburger Str. 35, D-01187 Dresden, Germany 
        \and Institute of Atmospheric Physics of the Czech Academy of Sciences, Prague, Czech Republic
        \and Space Research Institute, Austrian Academy of Sciences, Graz, Austria   
        \and Astronomical Institute of the Czech Academy of Sciences, Ondv{r}ejov, Czech Republic
        \and Space and Plasma Physics, School of Electrical Engineering and Computer Science, KTH Royal Institute of Technology, Stockholm, Sweden
}
 



\authorrunning{A. Vecchio et al.}

\date{\today}
\abstract 
{In order to allow for a comparison with the measurements from other antenna systems, the voltage power spectral density measured by {the Radio and Plasma waves receiver (RPW)} on board Solar Orbiter needs to be converted into physical quantities that depend on the intrinsic properties of the radiation itself (e.g., the brightness of the source).}
{The main goal of this study is to perform a calibration of the RPW dipole antenna system that allows for the conversion of the voltage power spectral density measured at the receiver's input into the incoming flux density. }
{We used space observations from the Thermal Noise Receiver (TNR) and the High Frequency Receiver (HFR) to perform the calibration of the RPW dipole antenna system. {Observations of type III bursts by the Wind spacecraft are used to obtain a reference radio flux density for cross-calibrating the RPW dipole antennas.
The analysis of a large sample of HFR observations (over about ten months), carried out jointly with an analysis of TNR-HFR data and prior to the antennas' deployment, allowed us to estimate the reference system noise of  the TNR-HFR receivers.}}
{ {We obtained the effective length, $l_{eff}$, of the RPW dipoles and the reference system noise of TNR-HFR in space, where the antennas and pre-amplifiers are embedded in the solar wind plasma. The obtained $l_{eff}$ values are in agreement with the simulation and measurements performed on the ground. } By investigating the radio flux intensities of 35 type III bursts simultaneously observed by Solar Orbiter and Wind, we found that while the scaling of the decay time as a function of the frequency is the same for the Waves and RPW instruments, their median values are higher for the former. This provides the first observational evidence that Type III radio waves still undergo density scattering, even when they propagate from  the source, in a medium with a plasma frequency that is well below their own emission frequency.}
{}

\maketitle

\section{Introduction} 
\label{sec:intro}
The Solar Orbiter (SO) mission \citep{Muller2020,Zouganelis_SAP} carries ten instruments, including the Radio and Plasma Wave experiment (RPW) \citep{RPW20}, which is designed to measure magnetic and electric fields, plasma wave spectra, and polarization properties, as well as the spacecraft (S/C) floating potential and solar radio emissions in the interplanetary medium. In this paper, we focus on the two high-frequency receivers of RPW:  Thermal Noise Receiver (TNR), producing electric and magnetic power spectral densities in the frequency range from 4 kHz to 1 MHz, and High Frequency Receiver (HFR), only providing electric power spectral densities in the range from 375 kHz over to 16 MHz. 
{The TNR-HFR has two input channels, which allows us to perform simultaneous measurements from two selectable sensors configurations. Among those, the monopole modes measure the potential difference between an antenna and the S/C ground, while the dipole modes measure the potential difference between two antennas. A mode for the measurement of magnetic field fluctuations at high frequency through the search coil  \citep[see][]{RPW20} has also been defined. A large variety of configurations, characterized by frequency range, temporal, and spectral resolutions, etc.,  are programmable in a series of operating modes optimized for specific analyses \citep{RPW20}}. A given TNR setup combines simultaneous measurements (dipoles or monopole, or both) from the two channels and computes autocorrelations and cross-correlations between the signals, allowing us to derive, with goniopolarimetry methods, the full polarization of the incoming radio waves. {The} HFR is a sweeping receiver providing the electric power {spectral} density for dipole antennas only. The three monopole sensors of TNR-HFR are named V1, V2, and V3 and they correspond to the antennas Pz, Py, and My, respectively \citep[see Figure 7 from][]{RPW20}.

The spectral properties of the signals measured by radio receivers on board the spacecraft are computed onboard. {The voltage power spectral density $V^2_\nu$ is usually sampled and transmitted on the ground}. The power spectral density needs to be converted to physical quantities related to the properties of the radiation itself in order to allow for a comparison with measurements performed by other antenna systems. To this purpose, instrumental parameters, such as the antennas' effective lengths and {capacitances}, need to be measured in flight in the real physical conditions of operation that (due to the effect  of the spacecraft body and the low density environment) are not reproducible on the ground.
A typical\ approach to performing this conversion is to observe a known source of electromagnetic radiation and to relate the measured $V^2_\nu$ to the flux of the source.  A dominant natural radio emission in the frequency range from 500 kHz to 10 MHz is the radio background radiation of the Galaxy, which it is commonly used as a calibration source for space antennas  \citep{Dulk01,Zarka04,Zaslavsky11}. 
Unfortunately, the TNR-HFR receiver  suffers a strong electro-magnetic (EM) contamination due to the central power distribution unit (PCDU) radiated by the Solar Panels, {at 120 kHz and harmonics}, and the reaction wheel electronic box, {at 80 kHz and harmonics} \citep{Maksimovic1st}. This makes the observation of the galactic background much more difficult as it is masked by the instrumental noise due to the platform. 

In this study, we used type III solar radio bursts as calibration sources for the RPW antenna system. Fast electrons beams, originating from magnetic reconnection sites of solar flares and escaping  through open magnetic field lines to the interplanetary space, interact with the local plasma and give rise to the type III radio bursts that {are commonly observed in the frequency range 10 kHz-1 GHz, and are} characterized by fast frequency drifts (around 100 MHz/s in the metric range). Type III bursts are among the strongest radio emissions routinely occurring in the heliosphere, reaching flux densities up to 10$^3$-10$^4$ higher than the galactic background and certainly well above the instrumental TNR-HFR background. This strong signal can be used to cross-calibrate the RPW antenna system by comparing the TNR-HFR data with simultaneous measurements of  type III radio bursts by the radio receiver band 1 (RAD1, 20-1040 kHz) on Wind/Waves \citep{Bougeret95}.

The paper is structured as follows. In Section \ref{formulas}, we present some general relations between the signal measured onboard by the radio receiver and the incoming electromagnetic radiation. {In Section \ref{cal_tyIII}, these relations are applied to cross-calibrate the dipole antennas of  RPW by using  type {III} emissions measured by the Wind spacecraft as reference sources. This allows us to derive the in-flight measurement of the effective length of the three dipoles (Section \ref{antenne}). The dipole antenna calibration was then applied to several type III emissions, occurring between July 2020 and January 2021, and a statistical analysis of the type III burst decay times, measured by both RPW and Waves, is presented in Section \ref{decad}.  Finally, the procedure to derive the system noise background of TNR-HFR {by comparing ten months of HFR observations with the expected Galaxy signal and by an analysis of TNR-HFR data before the antennas' deployment} is described in  Appendix \ref{gal_cal}.}

\section{Short antenna dipole calibration}
\label{formulas}
{In the short-dipole approximation}, when the wavelength $\lambda$ of the measured radiation is much larger than the physical length of the antenna, the latter experiences an homogeneous time-varying electric field and the conversion of the signal in V$^2$/Hz, measured as the receiver's input into incoming flux density,  can be  obtained analytically \citep{Dulk01,Zarka04,Zaslavsky11}. The RPW  antennas can be considered as short dipoles throughout much of the measured THR-HFR frequency range.
In this approximation, and for a unpolarized source,  it is possible to derive a simple equation relating the squared voltage fluctuations at a given frequency $\nu$, (in units of V$^2$/Hz), induced on a dipole antenna to the {apparent} brightness of the source $ {B_v}$ (in units W/m$^2$/Hz/sr):
\begin{equation}
V_\nu^2=\frac{l_{eff}^2}{2}Z_0\int_{\Omega_S}B_\nu(\Omega) sin^3\theta d\theta d\phi, \label{V2}
\end{equation}
where $\Omega_S$ is the solid angle  occupied by the source, $l_{eff}={V}/{E}$ ($V$ is the voltage induced at the antenna  by an incoming electromagnetic plane wave of electric field, $E$, when the electric field direction is parallel to the direction of the antenna \citep{Kraus03}) is the effective length of the antenna; a term $\sin^2\theta$ (where $\theta$ is the angle between the wave vector, $k,$ and the linear antenna direction) is due to the response pattern of a dipole antenna and $Z_0 = \sqrt{\mu_0/\epsilon_0} \approx 120\pi$ is the impedance of vacuum.
The antenna is electrically connected to a large input impedance receiver through the deployment mechanism (the antenna base) and various connection cables providing a so-called stray impedance $Z_s$. The effective voltage power spectral density $V^2_r$ measured by the receiver is then:
\begin{equation}
V_r^2=\left|\frac{Z_s}{Z_a+Z_s}\right|^2 V_\nu^2=\Gamma^2V_\nu^2
\label{Zs}
,\end{equation}
where $Z_a$ is the impedance of the antenna and $\Gamma$ is the gain factor. Both $Z_s$ and $Z_a$ are determined by the spacecraft design. In the radio frequency range (at frequencies well above the kHz), the resistive part of these impedances is negligible (typically, at 1 MHz, $1/\omega C_{a,s}\sim10$ $k\Omega$ and $R_{a,s} \sim 1$ $\Omega$) and the gain factor becomes: 
\begin{equation}
\Gamma = C_a/ (C_a + C_s),
\label{gain_fact}
\end{equation}
where $C_a$ and $C_s$ are the antenna and the stray capacitances, respectively. This indicates that the radio receiver's sensitivity can be increased by minimizing the stray capacitance.

{A general relation for the voltage power spectral density at the receiver level is \citep{Dulk01,Zarka04,Zaslavsky11}:
\begin{equation}
V^2_r=V^2_{noise}+V^2_{QTN}+V^2_{gal}+V^2_{source} \;\;\;.
\label{eq01}
\end{equation}
{{Equation (\ref{eq01}) relates the effective voltage power spectral density measured by the receiver $V^2_r$ to the signal from the source $V^2_{source}$, both expressed in $V^2/Hz$. Three other contributions have been included. Firstly $V^2_{noise}$ represents the noise produced by the electronic components of the receiver that superimposes any external signal and depends on both the receiver and  the impedance connected to it. The second contribution is due to the quasi-thermal noise (QTN) produced by the ambient plasma \citep{Zaslavsky11,Meyer17}. The third contribution is due to the Galactic signal.
By defining the power response of the antenna $P_\nu$ as
\begin{equation}
P_\nu= \frac{1}{2}\int_{\Omega_S}B_\nu(\Omega) sin^3\theta d\theta d\phi, 
\label{resp_ant}
\end{equation}
Equation (\ref{eq01}) becomes}:
\begin{equation}
V^2_r=V^2_{noise}+V^2_{QTN}+\Gamma^2{l_{eff}^2}Z_0 P^{gal}_\nu+\Gamma^2{l_{eff}^2}Z_0 P^{source}_{\nu} 
\label{eq1}
,\end{equation}
linking $V^2_r$  to the convolution of the brightness of the source with the antenna response function, $P^{source}_\nu$, expressed in W/m$^{2}$/Hz}.
If no external radio sources are present and  the Galaxy is the only external homogeneous  (i.e., $B_\nu$ does not depend on $\Omega$) source, which extends over $\Omega_s = 4\pi$, Equation (\ref{V2}) can be immediately integrated to yield:
\begin{equation}
V_\nu^2=\Gamma^2 l_{eff}^2 Z_0 P^{gal}_\nu=\frac{4\pi}{3}{\Gamma^2l_{eff}^2}Z_0 B^{gal}_\nu 
\label{Vgal}
\end{equation}
and Equation (\ref{eq1}) becomes
\begin{equation}
V^2_{r}=V^2_{noise}+V^2_{QTN}+\frac{4\pi}{3}\Gamma^2{l_{eff}^2}Z_0 B^{gal}_\nu
\label{eq1gal}
,\end{equation}
where the product  $\Gamma l_{eff}$ can be considered as a ``reduced effective length" of the antenna. 
\section{Calibration using solar type III bursts}
\label{cal_tyIII}
To evaluate the reduced effective length of the RPW electric antennas, we compared the radio spectra of a type III emission measured by RPW with the measurements by Wind/Waves when both spacecraft are close. On March 8, 2020, two type III emissions occurred  between 15:00-16:00 and 20:00-21:00, respectively. On March 30, 2020, a type III emission occurred between 11:00-12:00. These solar radio emissions were clearly observed by both Wind/Waves and SO/RPW (Figure \ref{dynamic_raw}).  
\begin{figure*}[]
\centering{
\hbox{
\includegraphics[scale=0.43]{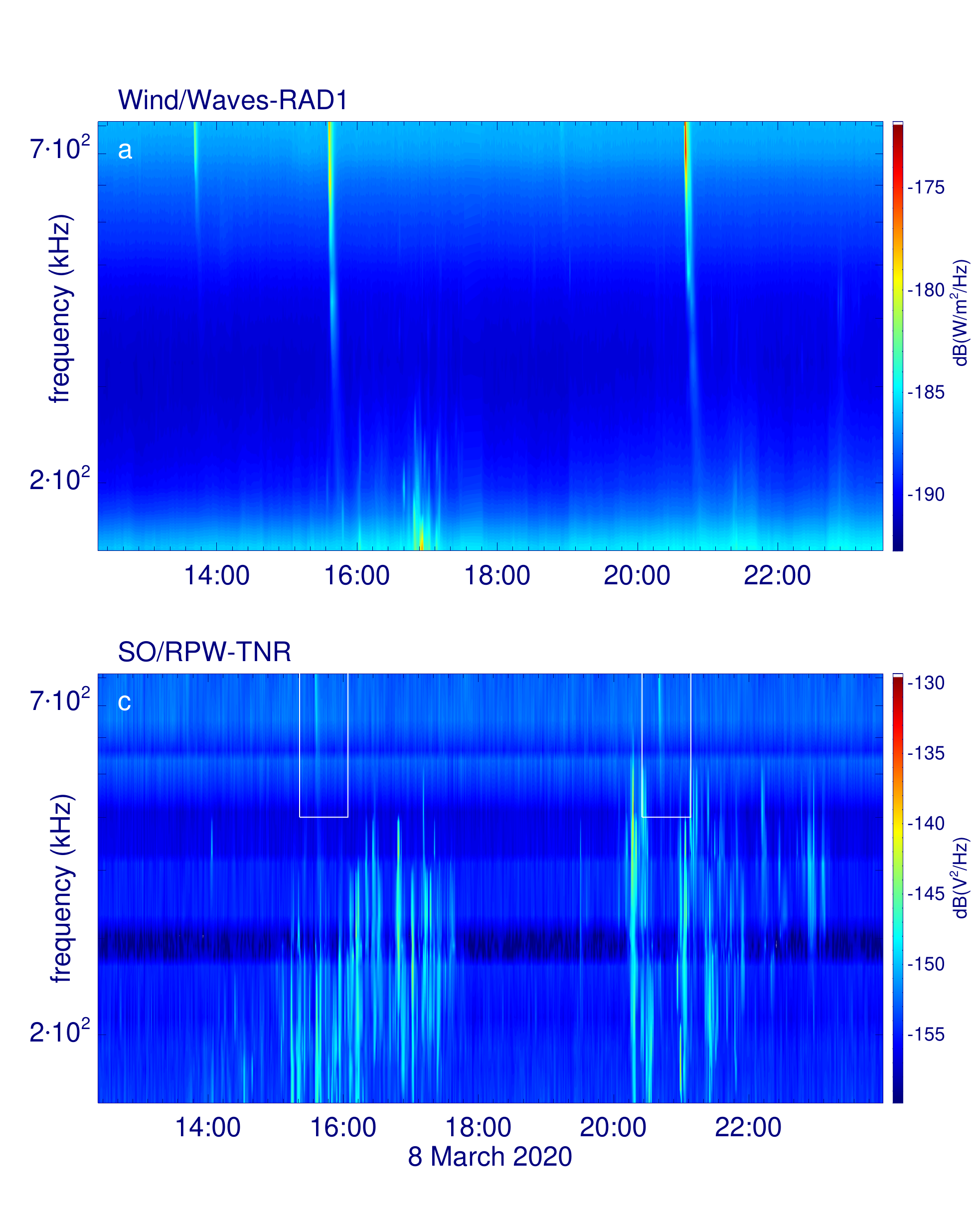}
\includegraphics[scale=0.43]{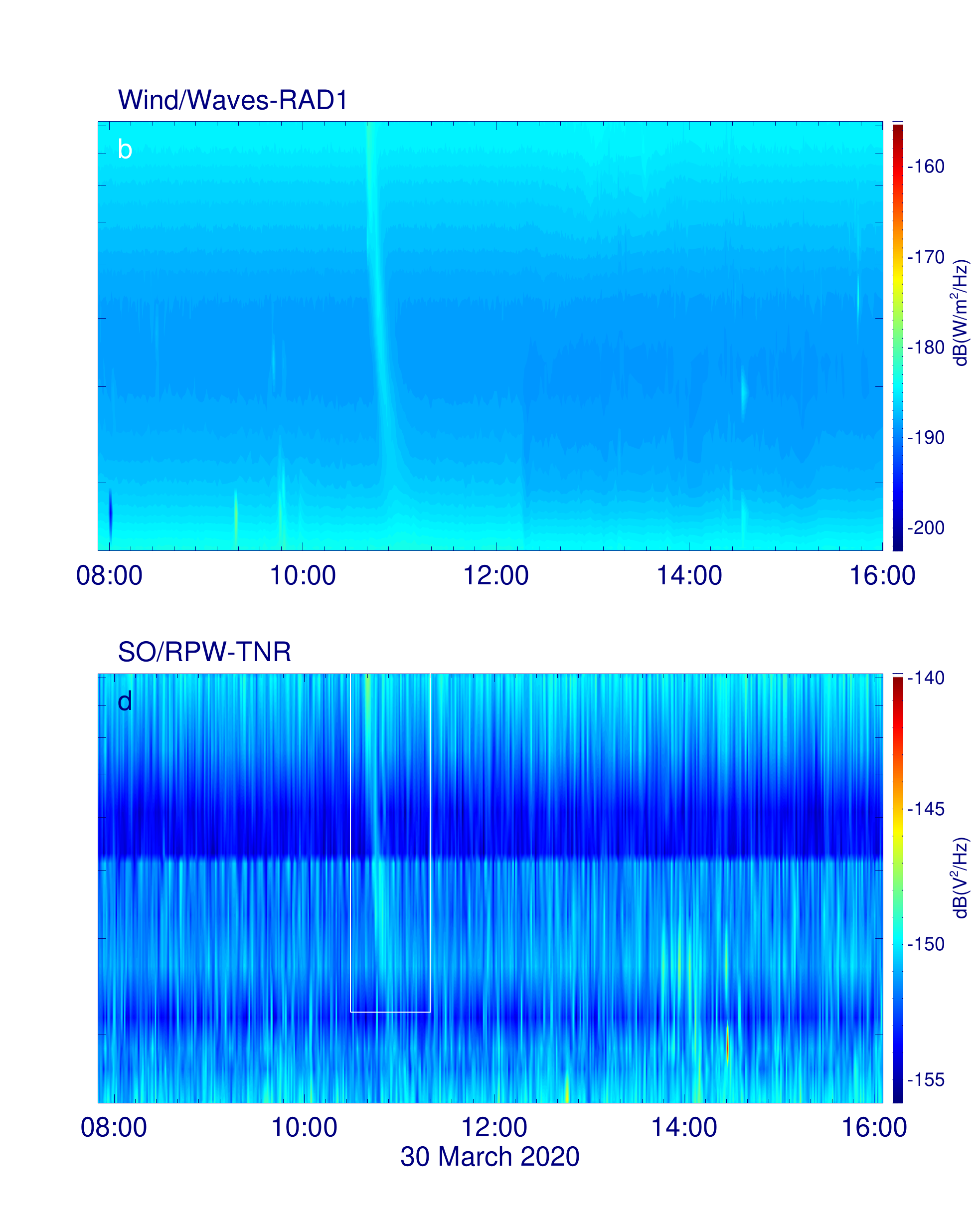}
}
}
\caption{Dynamic spectra from Wind/Waves-RAD1 (upper panels) and SO/RPW-TNR (dipole V1-V2, lower panels) for March 8 (a, c) and March 30, (b, d) 2020. White boxes underline the type III occurrence on the RPW data. The color tables are in dB, calculated from W$^2$/m$^2$/Hz for Wind and V$^2$/Hz for RPW.}
\label{dynamic_raw}
\end{figure*}
{We used the Wind/Waves flux density, obtained through goniopolarimetric inversions  \citep{Manning80} and expressed in W m$^{-2}$Hz$^{-1}$;} RPW data, on the other hand, were only calibrated at the receiver level and are expressed in V$^2$/Hz.
We used TNR-HFR dipole measurements that are affected to a lesser {extent}, with respect to the monopole ones, by the EM contamination, and we also chose clean frequencies. While the configuration of the HFR receiver was the same on March 8 and 30, with only the dipole V1-V2 measurements available, the configuration of TNR was different: over the first day, all the monopoles and only V1-V2 measurements were available, while all the monopoles and all the dipoles measurements were available for March 30. 
{Since the type III radio emissions are not isotropic and the exact position and  size of the source is unknown, we made use of the general Equation (\ref{eq1}) under some assumptions:
i) the two spacecraft are about at the same location, implying that they {observe the source with the same solid angle} and measure the same  flux density of the type III burst, $S_{B-RPW}(\nu)=S_{B-Wind}(\nu)$; ii) the k vector of the radiation is perpendicular to the RPW antennas.

Figure \ref{position}, showing the position of Wind and SO in the GSE reference frame, 
\begin{figure}[]
\centering{
\hbox{
\includegraphics[scale=0.2]{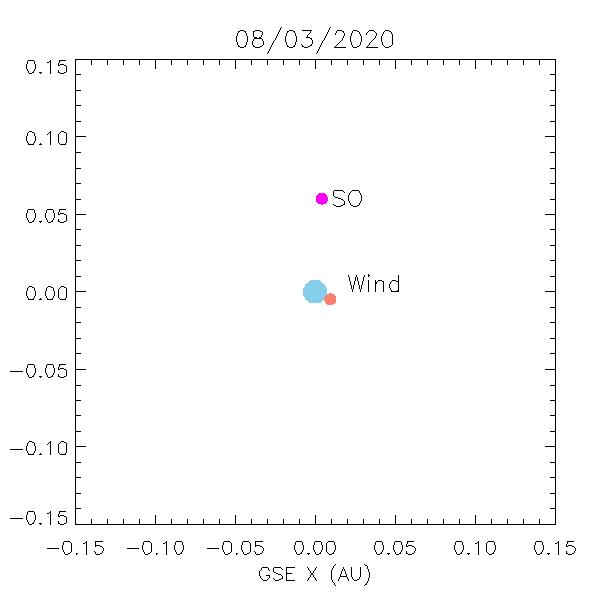}
\includegraphics[scale=0.2]{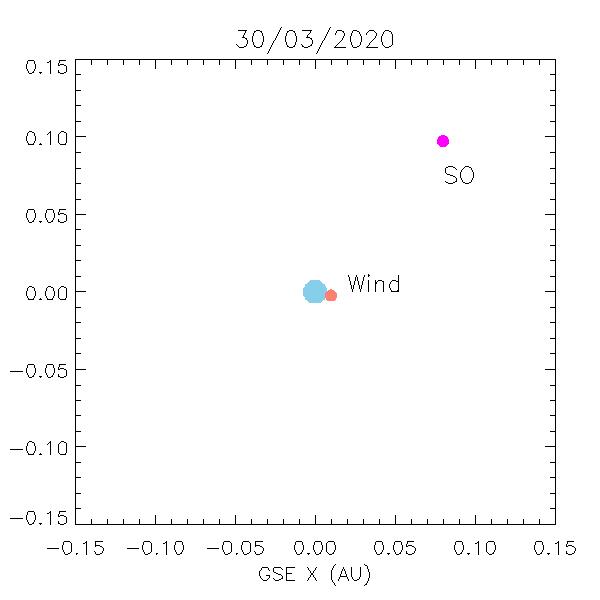}
}
}
\caption{Position of the SO (magenta dot) and Wind (orange dot) on March 8 (left panel) and  March 30 (right panel), 2020 in the GSE reference frame. The light blu dot indicates the Earth. {The X axis points towards the Sun.}}
\label{position}
\end{figure}
indicates that on both March 8 and 30, Wind and SO were very close each other (between 0.05 and 0.1 AU). {The frequencies at which the type III radiation is observed on both Wind and RPW (500 kHz-16MHz) are compatible with an emission occurring close to the Sun. Indeed, using the \citet{Leblanc98} density model, a radio emission emitted at a fundamental plasma frequency of 500 kHz would  be located within 10 solar radii from the Sun. The full width at half maximum (FWHM) of the type III radio beam at frequencies higher than 500 kHz is  about 100 degrees for both latitude and longitude \citep{Bonnin08}. On the other hand, the angular distance between the two spacecraft, as viewed from the source, is 3-5 and 0.75-1 degrees in longitude and latitude respectively. Since the angular width of the radio beam is very large with respect to the mutual distance between Wind and SO, the difference in the radio flux measured by the two spacecraft is low. When quantified through  the functional form of a type III radio beam derived by \citet {Bonnin08}, the difference in radio flux is about 0.3\% and 0.1\% for the longitude and latitude, respectively. At the same time, the difference in radio flux  due to the different distances of the two spacecraft from the Sun is {about } 3\%. {Assumption i) can be thus considered, to a good approximation, as valid}.
Concerning assumption ii), since the radio emission would be located close the Sun, the k vector of the radiation can be considered, {to a good approximation,} as perpendicular to the RPW antennas since the latter are perpendicular to the Sun-spacecraft direction. 
Under these assumptions, the relation between the RPW voltage power spectral density and the flux density during the type III burst can be expressed as:
\begin{equation}
V^2_{B-RPW}(\nu)= (1/2) Z_0\Gamma^2 l_{eff}^2 S_{B-RPW}(\nu) \;\; .
\label{VS}
\end{equation}
The factor $\Gamma l_{eff}$ can be thus derived from Equation (\ref{VS})  by comparing Waves and RPW data at the frequencies showing the type III emission: 
\begin{equation}
\Gamma^2 l^2_{eff}=\frac{2}{Z_0}\frac{V^{2}_{B-RPW}(\nu)}{S_{B-Wind}(\nu)} 
\label{gammal}
,\end{equation}
where $V^{2}_{B-RPW}$ and $S_{B-Wind}$ are the power spectral density and the flux density as measured by RPW and Waves, respectively. Then,}
$V^{2}_{B-RPW}$  follows from Equation (\ref{eq01}) 
\begin{equation}
V^{2}_{B-RPW}(\nu) = V^{2}_{r}(\nu) -V^{2}_{gal}(\nu)-V^{2}_{noise}(\nu)\;\;\;\;\;\;\;\;\;\;\;\;V^{2}/Hz 
\label{vb}
,\end{equation}
where $V^{2}_{r}$ is the power spectral density as measured by TNR and HFR. The term $V^{2}_{gal}+V^{2}_{noise}$ has been estimated as the median value of the power spectral density calculated in time intervals far from the type III occurrence. We remark that the results do not change if the term $V^{2}_{gal}+V^{2}_{noise}$ is evaluated by using, for instance,{} the lower 5\% percentile.
The contribution of the QTN can be considered negligible for $\omega \gg 500$ kHz. However, possible residual contributions of QTN are removed when the median background is subtracted {from} the data.
 \\
$S_B$ is derived from RAD1 measurements as:
\begin{equation}
S_{B-Wind}(\nu) = S_{Wind}(\nu) -S_{backg-Wind}(\nu)\;\;\;\;\;\;\;\;\;\;\;\;W/m^{2}/Hz
\label{sb}
,\end{equation}
where $S_{Wind}$ is the type III flux density as measured by Waves and $S_{backg-Wind}$ represents the background signal due  to other sources of noise at low frequency (e.g., the galactic background noise). As before, $S_{backg-Wind}$ is estimated as a median value and subtracted {from}  the Waves spectra  to remove the low frequency noise from the data.  Moreover, since RAD1 and TNR/HFR produce spectra at slightly different frequencies and are characterized by different time sampling, we firstly interpolated $S_{B-Wind}$ on the same time grid of TNR and HFR data and then by applying Equation (\ref{gammal}), we choose $S_{B-Wind}(\nu)$ at the closest frequency to the corresponding $V^{2}_{B-RPW}(\nu)$.
Examples of the spectra from RAD1, TNR, and HFR, during the type  III bursts of March 8, are shown in figure \ref{compare_sp}. 
\begin{figure*}[!ht]
\centering{
\includegraphics[scale=0.4]{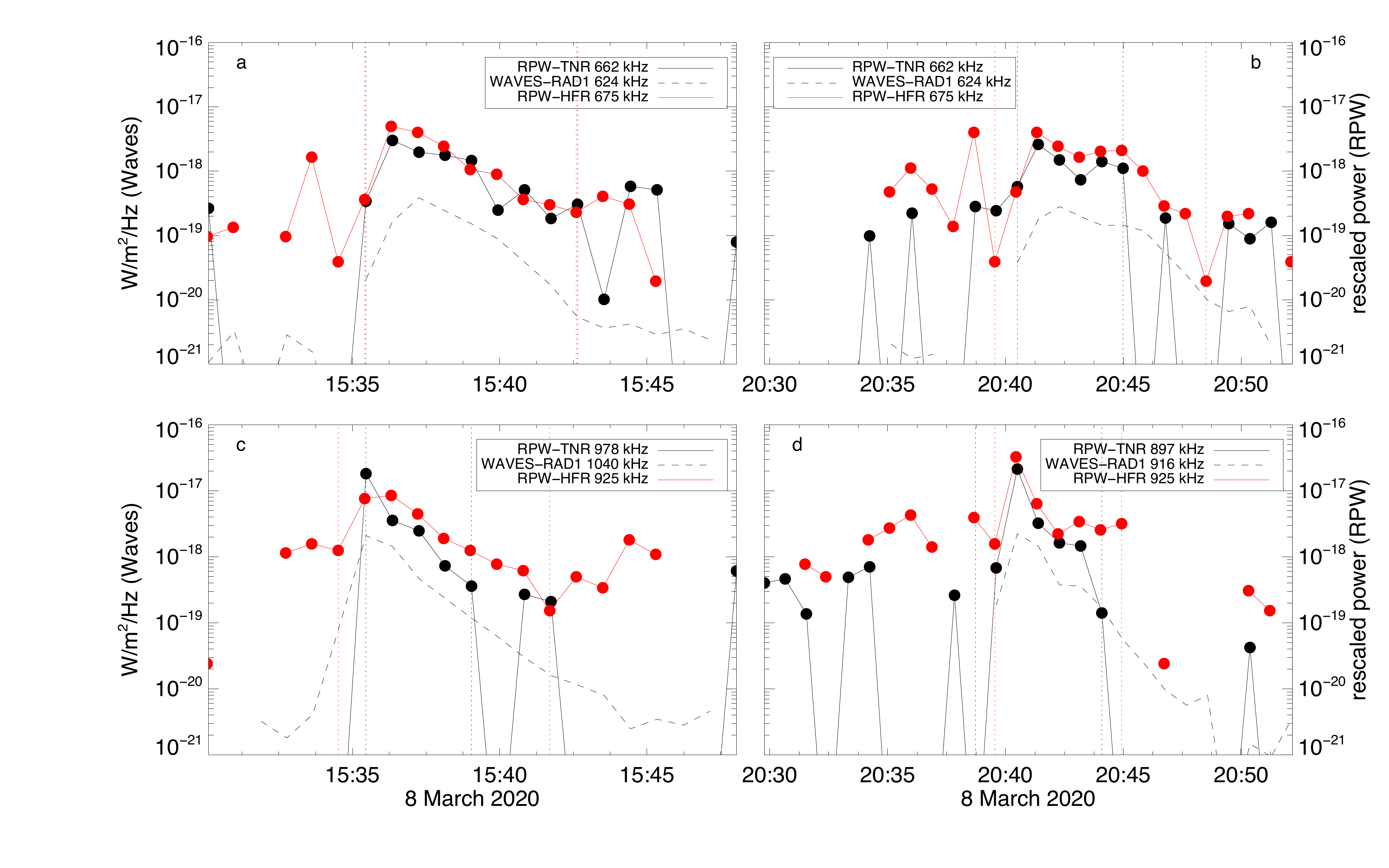}
}
\caption{Comparison between $S_{B-Wind}$ (dashed lines) in W m$^{-2}$Hz$^{-1}$, as obtained by Waves-RAD1, and $V^{2}_{B-RPW}$ (solid lines), as measured by RPW and rescaled by a factor ${2}/Z_0$  for RPW-TNR (black) and RPW-HFR (red) and for the V1-V2 dipole. Vertical dotted lines identify the data points used to calculate $\Gamma l_{eff}$. Panels a, c, and b, d refer to the two type III emissions detected on March 8, 2020, respectively. }
\label{compare_sp}
\end{figure*}
For each frequency, a value for $\Gamma l_{eff}$ is calculated by averaging Equation (\ref{gammal}) over the type III emission profile. The number of points in the average is not the same at each frequency since the duration of the type III emission profile changes as a function of the frequency.\\
\begin{figure}
\centering{
\includegraphics[scale=0.4]{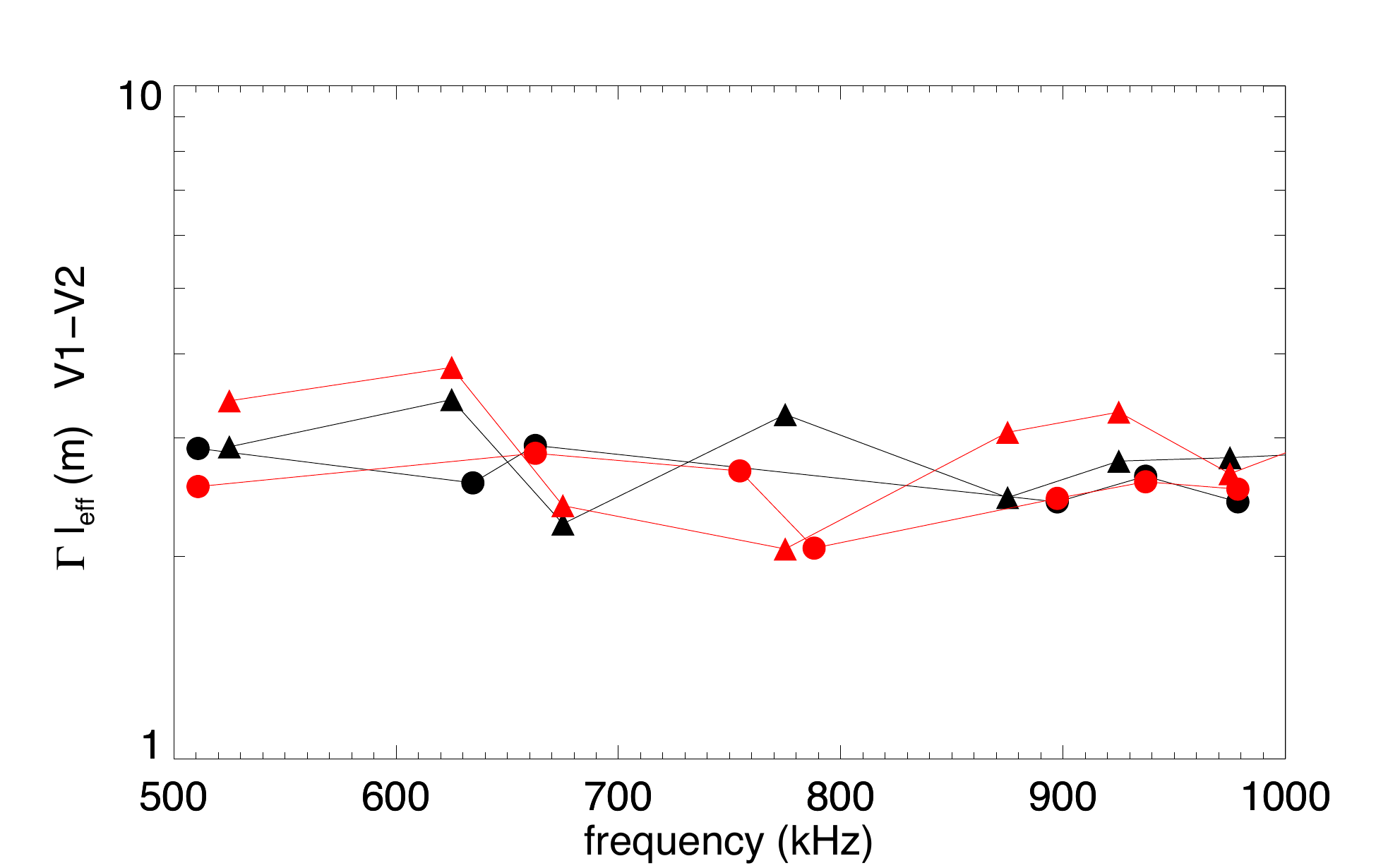}
}
\caption{$\Gamma l_{eff}$ as a function of the frequency for the first (black) and second (red) type III on March 8th 2020. Dots and triangles refers to TNR and HFR, measurements respectively.
}
\label{ratio_freq}
\end{figure}
For both days, an average $\Gamma l_{eff}$ for V1-V2  was calculated as a function of frequency from both the TNR and HFR data. The $\Gamma l_{eff}$ values for V2-V3 and V3-V1 were evaluated from the measurements on March 30 based only on TNR. An example of the average $\Gamma l_{eff}$  as a function of  frequency is shown in Figure \ref{ratio_freq} for both TNR and HFR measurements. This quantity should not depend on the frequency since neither the effective length nor the antenna capacitance does so in the frequency range where the short dipole approximation holds. Indeed, no strong variations were detected for the same type III nor when the samples were compared. For each day, we calculated a single $\Gamma l_{eff}$ as the average over frequency, receiver (if measurements on both TNR and HFR are available) and type III (for March 8 data when two emissions are detected). The average values of $\Gamma l_{eff}$ are shown in table \ref{tab0}.
\begin{table}[h]
\caption{ {Average $\Gamma l_{eff}$ values as obtained from the cross-calibration with type III measurements from Wind/Waves on March 8 and 30, and from the Galaxy background signal from ten months of HFR V1-V2 dipole data (see Appendix \ref{gal_cal}).}}
\label{tab0}
\centering
\begin{tabular}{cccc}
\hline
 { Sensor} &{March 8 } & { March 30} & { Galaxy}  \\ \hline\hline
 V1-V2  & $2.7\pm0.3$ m & $3.2\pm0.3$ m& $3.4\pm0.1$ m  \\
V2-V3  &  &$2.5\pm 0.3$ m \\
V3-V1 & &$3.0\pm 0.2$ m\\
\hline
\end{tabular}
\end{table}
The comparison between V1-V2 $\Gamma l_{eff}$'s  from the two days indicates that the values are compatible within the uncertainty limits. {The $\Gamma l_{eff}$ value in the last column of table \ref{tab0} is derived by comparing about ten months of V1-V2 HFR data, to reduce the EM contamination, with a model of the Galaxy background (see Appendix \ref{gal_cal}). Only HFR measurements for the V1-V2 dipole were available for such a long period. The value of $\Gamma l_{eff}=3.4\pm0.1$ m is compatible, in the uncertainty limit, with the values obtained from the type III emissions. We also remark that the $\Gamma l_{eff}$ values are compatible with the corresponding ones obtained after ground testing and characterizations of the RPW antenna expected to be between $2.35$ and $3.24$ m \citep{RPW20}.}
Unfortunately, no observations of type III emissions by TNR monopoles are available for the periods when SO was close to Wind. The calibration of the RPW monopole antennas is thus ongoing and will be the subject of a future paper.
\section{Determination of the effective length}
\label{antenne}
Given that for March 30, the measurements for all the three dipoles are available, in the following we  use $\Gamma l_{eff}$ values obtained from this day as a reference for calculating the effective length, $l_{eff}$, of the SO dipole antenna.
To derive $l_{eff}$,  we first need to  model the gain factor, $\Gamma,$ and to this end, we need to know the antenna capacitance, $C_a$, and the stray capacitance, $C_s$. The antenna capacitance can be analytically derived by considering cylindrical antennas of length $L,$  and radius $a$  \citep{Zaslavsky11}:
\begin{equation}
C_a=\frac{\epsilon_0\pi}{k}\frac{\tan{(kL)}}{\log{(L/a)}-1} \;\;\; .
\label{ant_capa}
\end{equation}
Values of $a=0.015$ m, $C_s=54.7$ pF, for the RPW dipoles,  $L=7.857$ m, for the dipoles 12 and 31, and $L=6.99$ m, for the dipole 23,  as given in Maksimovic et al. (2021) were used. We note that the latter paper provides corrected values for some parameters that were previously published in \citet{RPW20}. By using these parameters,  $l_{eff}$ can be obtained from the previously derived values for  $\Gamma l_{eff}$ . The effective antenna lengths for the three TNR-HFR dipoles are shown in Table \ref{tab1}. The value of {$l_{eff}$ obtained by using the $\Gamma l_{eff}$ as derived from the Galaxy background signal from ten months of HFR measurements is also shown}. The values of $l_{eff}$ agree, within the uncertainty limits, with the values of the  effective length for the dipoles found by \citet{Panchenko17} and reported by \citet{Maksimovic19}.
\begin{table*}
\caption{Results of the Calibration. {Parameters of the RPW dipole antennas as obtained from the cross-calibration with type III measurements from Wind/Waves, from the Galaxy background signal from ten months of HFR V1-V2 dipole data (see Appendix \ref{gal_cal}) and from numerical simulations by \citet{Panchenko17}.}}
\label{tab1}
\begin{tabular}{ccccc}
\hline
{Antenna Configuration} &{average $\Gamma$} & { $l_{eff}$ (m) typeIII} & { $l_{eff}$ (m) Galaxy} & { theoretical $l_{eff}$ (m) }   \\ \hline\hline
 V1-V2 & 0.44$\pm$0.01 & 7.2$\pm0.8$ & 7.7$\pm$0.3 & $7.53$ \\
V2-V3 & 0.42$\pm$0.01 & 6.0$\pm0.8$ & &$5.60$ \\
V3-V1 & 0.44$\pm$0.01 & 7.0$\pm0.6$ & &$ 7.53$\\
 \hline
\end{tabular}
\end{table*}

\section{Type III radio bursts simultaneously measured by SO and Wind}
\label{decad}
This section deals with the statistical analysis and comparison between type III bursts simultaneously observed by SO and Wind. We apply the antenna calibration developed in the previous section to a selected set of 79 type III radio bursts observed by SO between June 2020  and January 2021, representing the most intense and most simple cases detected.  
Due to the strong EM pollution suffered by RPW measurements, not all frequency channels are capable of detecting the type III signal. Since the purpose of the analysis is to compare Wind and SO measurements  for the type III detected on both spacecraft only, a further selection was made according to the following steps. First, we identified the TNR frequency channels where the type III burst peak is above the background plus $4\sigma_{back}$ (where $\sigma_{back}$ is a measure of amplitude  of the fluctuations calculated as the standard deviation in the same time interval where the background is evaluated), then we verified that the same event is evenly detected at the RAD1 channel closer to the corresponding TNR frequency.  {If both these requirements were verified, the event {was} selected for subsequent analyses. The final database is then reduced to 35 type III events.}

\begin{figure}[ht!]
\centering{
\includegraphics[scale=0.3]{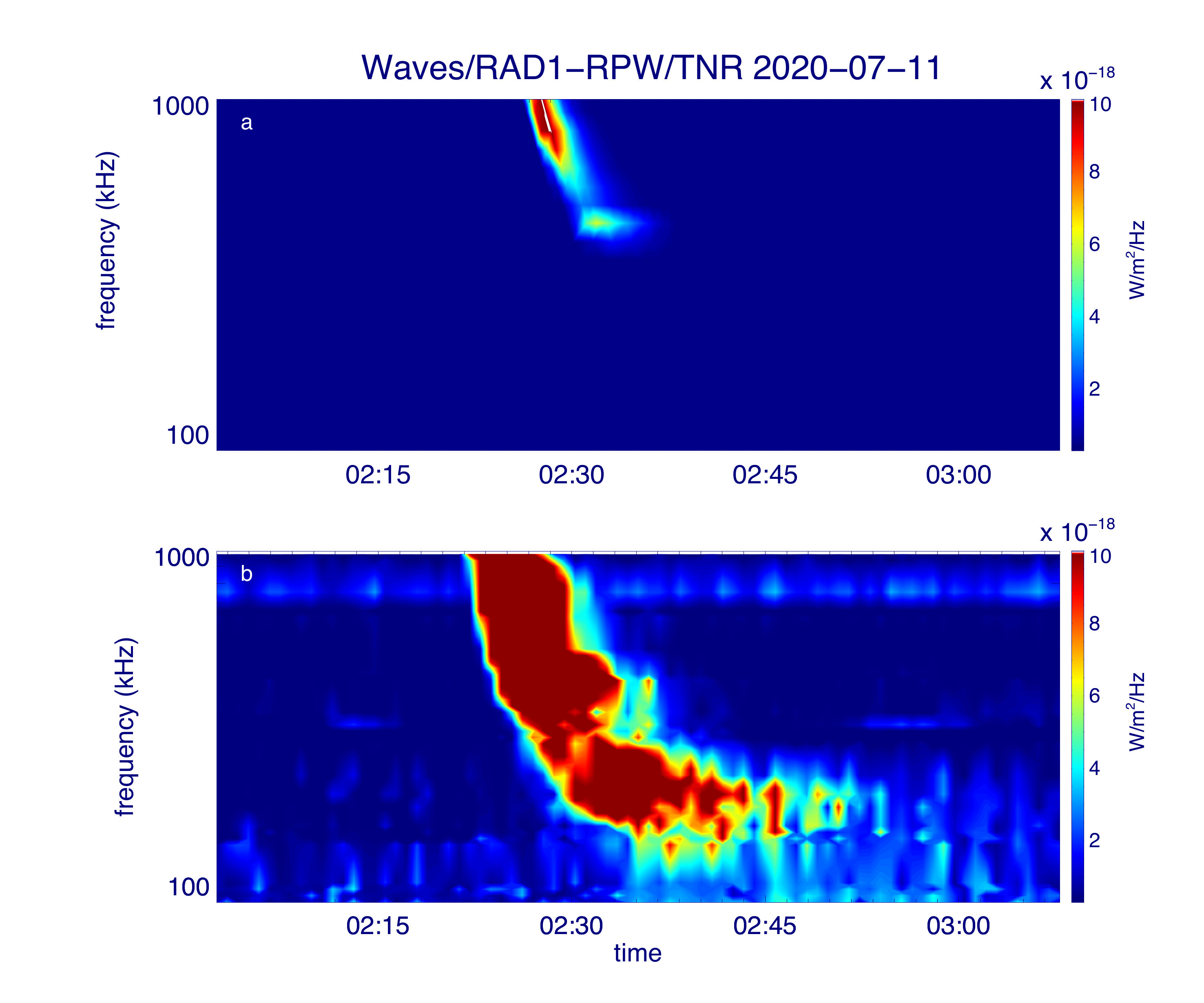}
}
\caption{Radio measurements of the 2020 July 11 type III burst. {Radio flux density  for Wind/Waves-RAD1 (a) and SO/RPW-TNR (b).}}
\label{typ_exe}
\end{figure}
One example from one of the considered events is shown in Figure \ref{typ_exe}. It is a type III burst occurring on 2020 July 11 that has been linked to an electron beam at $\sim$ 200 keV  detected by the EPD instrument \citep{Pacheco20} onboard SO at around 02:30. Figure \ref{typ_exe} shows the flux density from Wind/Waves-RAD1 (a) and SO/RPW-TNR (b). Both spacecraft detected a simple and isolated type III burst with an onset time at about 02:22 UT. During this event, the GSE {coordinates of the two spacecraft were $(x_{Wind},y_{Wind})=(0.01,0.004)$ AU and $(x_{SO},y_{SO})=(1.19,0.58)$ AU and the distance from the Sun was $0.99$ AU and $0.88$ AU for Wind and SO respectively.}
\begin{figure*}[ht!]
\centering{
\includegraphics[scale=0.6]{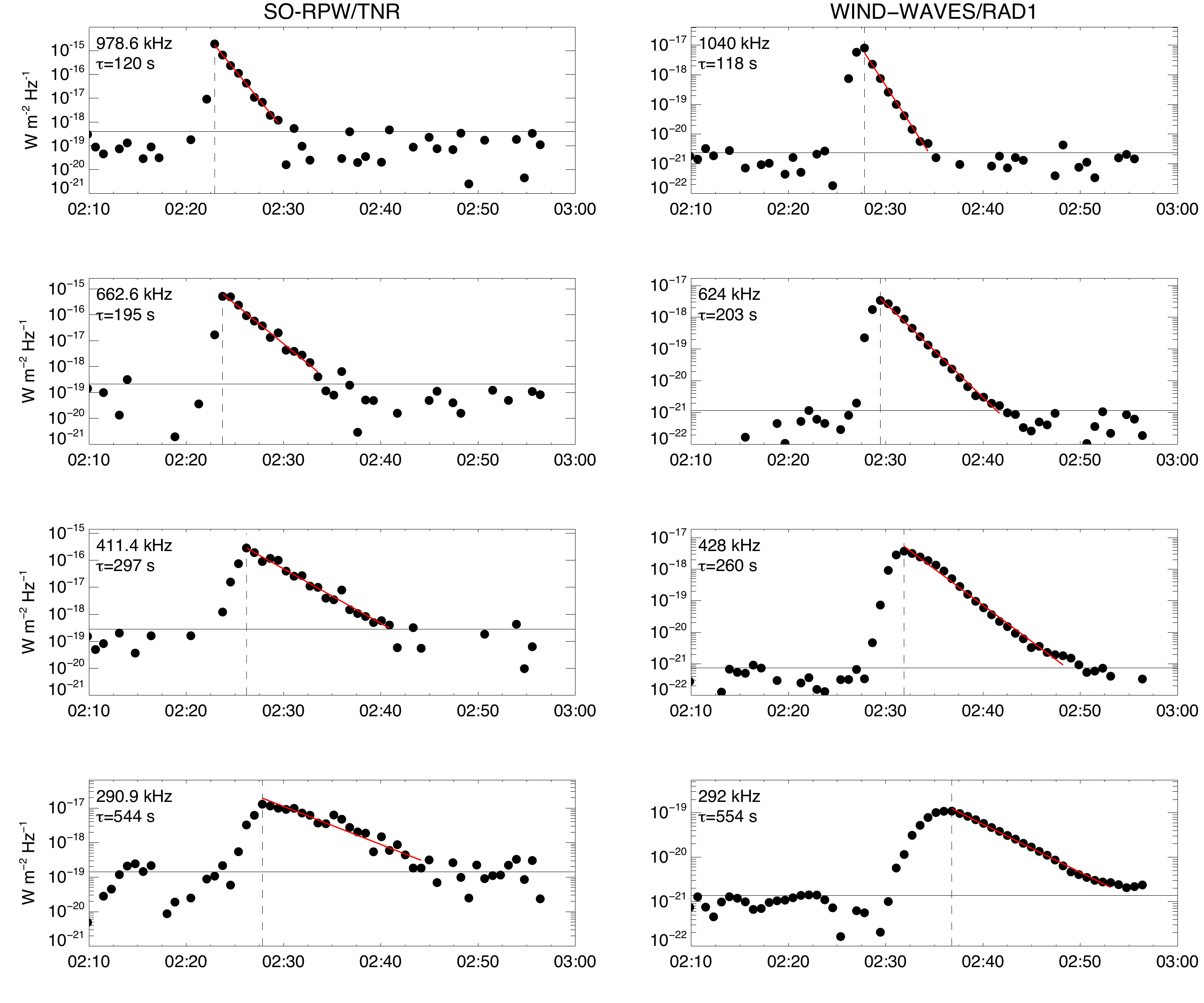}
}
\caption{{Radio measurements of the 2020 July 11 type III burst. Light curves of the radio flux density measured by SOLO-RPW/TNR (left column) and Wind-Waves/RAD1 (right column) at four frequency channels. The red line shows the results of decay time fitting. The dashed vertical line indicates the peak fluxes  and the horizontal line shows the median value.}
}
\label{light_c}
\end{figure*}
The time profile of the intensity of the type III at a given frequency consists of a very fast rising phase, followed by a longer phase decreasing exponentially \citep{Evans73}. This is clearly shown in Figure \ref{light_c}, where the light curves at four frequencies $[290.9 ,411.4, 662.6, 978.6]$ kHz for RPW and $[292,428,624,1040]$ kHz for Waves are plotted. Results from \citet{Krupar18} suggest that the characteristic exponential decay profile of type III bursts could be solely explained by the scattering of the radio beam by electron density inhomogeneities as they propagate from the source to the spacecraft. 

To evaluate the characteristic decay time $\tau$ for each type III burst, the background level was  first calculated as the median value, far from the type III occurrence, of the flux density S for each frequency channel. Then the data points between the peak time, $t_{peak}$, and the last value above the background were fitted through an exponential function of the form:
\begin{equation}
S(t)=S_{peak}\exp\left(\frac{t_{peak}-t}{\tau}\right) \;\;.
\label{eq_decay}
\end{equation}
An example of the result of the fitting procedure is shown in Figure \ref{light_c}.  As expected, the calculated decay times $\tau$ increases with decreasing frequency. 
\begin{figure*}[ht!]
\centering{
\includegraphics[scale=0.6]{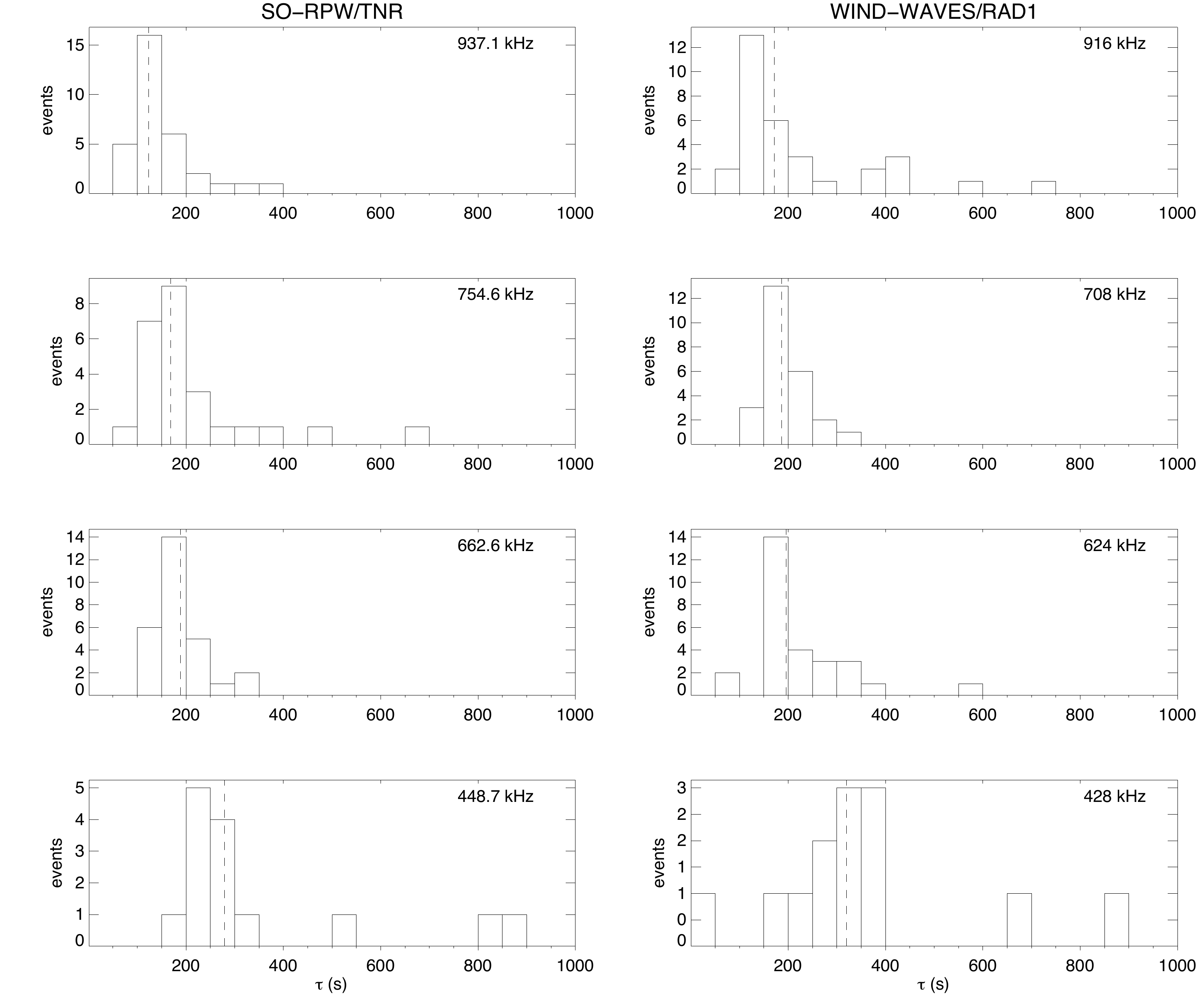}
}
\caption{Histogram of the $\tau$ calculated for each type III of the dataset at different frequencies for SOLO-RPW/TNR (left column) and Wind-Waves/RAD1 (right column). Dashed line indicates the median value.
}
\label{histo_tyIII}
\end{figure*}
The decay times $\tau$ have been evaluated at different frequencies for all the  type III events analyzed. Histograms of the decay time for four frequency channels are shown in Figure \ref{histo_tyIII}. The statistics on the considered events confirm that, on average, the values of $\tau$ increase with decreasing frequency. The number of detected events also decreases at lower frequencies. The latter effect, already observed in other datasets \citep{Krupar14,Krupar18,Krupar20}, may be due to the combination of the intrinsic properties of the radiation mechanism, the effect of the directivity of the radiation, and propagation effects between the source and the observer \citep{Leblanc95,Musset21}. 
Figure \ref{med_decay} shows the median values of decay times as a function of frequency for both SO and Wind dataset. As already performed in previous works \citep[e.g.,][]{Evans73,Alvarez73,Krupar18,Krupar20}, 
a decay time dependence on frequency in the form of a power law,
\begin{equation}
\tau(\nu)=\alpha\nu^{\beta}
\label{powlaw}
,\end{equation}
 is considered here.
\begin{figure}[ht!]
\centering{
\includegraphics[scale=0.3]{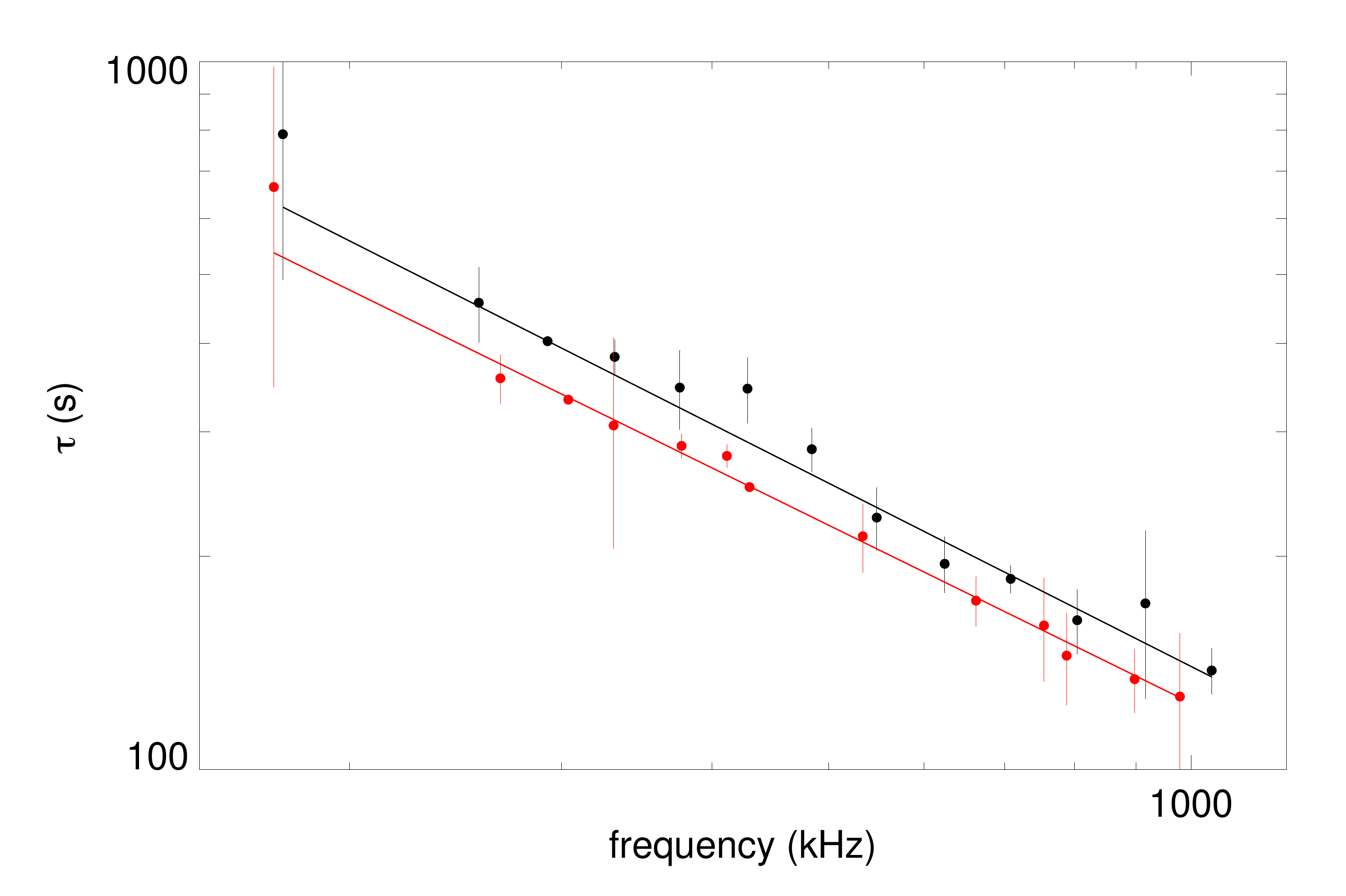}
}
\caption{Median values of $\tau$ for SO (red) and Wind (black) as a function of frequency. Error bars represent the 25th and 75th percentiles. The black and red lines represent the results of power-law fitting from Equation  (\ref{powlaw}).}
\label{med_decay}
\end{figure}
The spectral indices $\beta$, obtained by fitting SO and Wind data through the function in Equation (\ref{powlaw}) in the frequency range $\sim170$ kHz - $\sim1$ MHz, are equal to $-0.83\pm0.4$ for SO and $-0.86\pm0.3$ for Wind  and compatible in the uncertainty limits between each other. These values can be compared to the results of other works present in the literature. The spectral index $\beta$ found by \cite{Alvarez73}, in the  frequency range 50 kHz-3.5 MHz, from data by the spacecraft OGO-5, was $-0.95$. \citet{Evans73} found a value of $\beta=-1.08$, between 67 kHz and 2.8 MHz, from data by  RAE-1 and IMP-6. More recently, \citet{Krupar18,Krupar20}, by performing the analysis of the decay time on a dataset of 152 type III bursts from STEREO-A/STEREO-B (in the range $0.1-1$ MHz) and 30 events for Parker Solar Probe (PSP) (in the range $\sim0.4-10$ MHz) spacecraft, found $\beta=-1.18\pm0.02$/$\-1.25\pm0.02$ (for STEREO A/B) and  $\beta=-0.60\pm0.01$ (for PSP). The values we obtained for SO and Wind are slightly higher with respect to the STEREOA/STEREOB results and comparable, given the error bars on our beta parameters, with PSP. The value we found for Wind is also higher than the $\beta=-1.81\pm0.22$ derived by \citet{Krupar20} from only one type III burst detected by Wind (although the fit is performed on a slightly different frequency range). 
{The differences among $\beta$ values from different studies can be attributed to several factors: data quality and time resolution, type III event selection, decay time computation, frequency band, etc. In comparison with the present study, differences can arise due to the list of frequency and the frequency range of the fit that differs from previous studies  thus changing the relative fitting weights allocated to frequencies.}
This could explain why lower or higher $\beta$ values, computed on lower or higher frequency bands, were found for STEREO/PSP.

The median values of decay times as a function of frequency are shown in Figure \ref{med_decay}. The $\tau$ values derived from the RPW dataset are always lower than the corresponding Wind values by about 45 seconds, on average. Considering that one of the possible physical mechanisms behind the exponential decay observed in type III profiles is believed to be the radio-wave scattering, this result could indicate that an extra-scattering occurs for radio waves reaching the Wind spacecraft at 1 AU with respect to RPW. Although the radio scattering is certainly very important close to the source, as previously reported \citep[see e.g.][]{Krupar18,Kontar19,Krupar20}, our observational results demonstrate that {it} still continues to affect type III radio burst propagation. \\
This observation represents the first evidence of a statistical difference in decay times measured by spacecraft located at different distances from the Sun. In a previous analysis, where only one type III burst measured by both   PSP and STEREO-A was considered, no significant difference in $\tau$ values from the two spacecraft in the overlapping frequency range were found \citep{Krupar20}. However our results -- obtained  by considering  a larger number of events  and performing a punctual comparison between Wind and SO events in a way that  only the same  type IIIs detected by both spacecraft were taken into account and statistically compared -- provide a more robust result.
Moreover,  while the number of events we considered in our analysis is not particularly large, we analyzed observations in a time period when  SO and Wind were always separated by a very high radial distance (greater than 1 AU), allowing the effect of the additional scattering to  be highlighted.
\section{Conclusions}
Applying the calibration method developed by \citet{Zaslavsky11} for the  STEREO/Waves instrument {we performed the antenna calibration of the RPW instrument} on the SO spacecraft. Although  the original method implied the use of the galactic background signal as a reference source,  the same approach is not entirely suitable for RPW due to the high level of EM pollution suffered. {By cross-calibrating the TNR-HFR data using type III burst emissions measured by Wind/Waves as reference sources, we provide accurate values of the reduced effective length of the dipole RPW antennas. The analysis of about ten months of HFR measurement and TNR-HFR data before the antennas' deployment, described in Appendix  \ref{gal_cal}, allowed us to derive the reference system noise of TNR-HFR in space, where the antennas and pre-amplifiers are embedded in the solar wind plasma. The reference system noise for both TNR and HFR receivers, as a function of frequency, is shown in Figure \ref{backg12}.  An almost constant background of about $-160$ dB is found for TNR in the frequency range 30 kHz-1 MHz. The TNR background level increase, observed at low frequencies, may be attributed either to noise generated by the spacecraft subsystems or to external noise picked up even in the absence of deployed antennas. The HFR background is almost constant, with an average value of $-157$ dB in the range 375 kHz- 4.3 MHz, while it increases at higher frequencies due to the effect of antenna resonances and due to possible couplings with the spacecraft.} 

By analyzing a sample of 35 type III radio bursts simultaneously observed  by both SO and Wind in the period between June 2020 and January 2021, we performed a statistical study on the retrieved  decay times $\tau$. While our results confirm some expected properties, such as the number of detected events decreasing at lower frequencies and the values of $\tau$ increasing with decreasing frequency according to a power law, quite interestingly we found that the  $\tau$ values from the RPW events are systematically lower than the corresponding Wind values by about 45 seconds, on average. This result, representing the first evidence of a statistical difference in decay times measured by spacecraft at different distances from the Sun, could indicate that the radio-wave scattering can still play a relevant role far from the source as well. Of course, future analyses with multi-spacecraft observations at various radial distances from the Sun (including future data from RPW), {together with quantitative comparisons with the results of numerical simulation describing the effects refraction and scattering on the propagation of interplanetary radio emissions},  will be performed to increase the statistical set and further asses the result obtained here.


\begin{acknowledgements}
{We acknowledge the anonymous Reviewer for the useful comments.} \\
Solar Orbiter is a space mission of international collaboration between ESA and NASA, operated by ESA. 
The Wind/WAVES investigation is a collaboration of the Observatoire de Paris, NASA/GSFC, and the University of Minnesota. 
V. Krupar acknowledges the support by NASA under grants \texttt{18-2HSWO218\_2-0010} and \texttt{19-HSR-19\_2-0143}. 
\end{acknowledgements}


\bibliographystyle{aa} 
\bibliography{references} 

\begin{thebibliography}{25}
\expandafter\ifx\csname natexlab\endcsname\relax\def\natexlab#1{#1}\fi

\bibitem[{{Alvarez} \& {Haddock}(1973)}]{Alvarez73}
{Alvarez}, H. \& {Haddock}, F.~T. 1973, \solphys, 30, 175

\bibitem[{{Bonnin} {et~al.}(2008){Bonnin}, {Hoang}, \& {Maksimovic}}]{Bonnin08}
{Bonnin}, X., {Hoang}, S., \& {Maksimovic}, M. 2008, \aap, 489, 419

\bibitem[{{Bougeret} {et~al.}(1995){Bougeret}, {Kaiser}, {Kellogg}, {Manning},
  {Goetz}, {Monson}, {Monge}, {Friel}, {Meetre}, {Perche}, {Sitruk}, \&
  {Hoang}}]{Bougeret95}
{Bougeret}, J.~L., {Kaiser}, M.~L., {Kellogg}, P.~J., {et~al.} 1995, \ssr, 71,
  231

\bibitem[{{Dulk} {et~al.}(2001){Dulk}, {Erickson}, {Manning}, \&
  {Bougeret}}]{Dulk01}
{Dulk}, G.~A., {Erickson}, W.~C., {Manning}, R., \& {Bougeret}, J.~L. 2001,
  \aap, 365, 294

\bibitem[{{Evans} {et~al.}(1973){Evans}, {Fainberg}, \& {Stone}}]{Evans73}
{Evans}, L.~G., {Fainberg}, J., \& {Stone}, R.~G. 1973, \solphys, 31, 501

\bibitem[{{Kontar} {et~al.}(2019){Kontar}, {Chen}, {Chrysaphi}, {Jeffrey},
  {Emslie}, {Krupar}, {Maksimovic}, {Gordovskyy}, \& {Browning}}]{Kontar19}
{Kontar}, E.~P., {Chen}, X., {Chrysaphi}, N., {et~al.} 2019, \apj, 884, 122

\bibitem[{{Kraus} \& {Marhefka}(2003)}]{Kraus03}
{Kraus}, J.~D. \& {Marhefka}, R.~J. 2003, {Antennas for All Applications} (New
  York: McGraw-Hill)

\bibitem[{{Krupar} {et~al.}(2018){Krupar}, {Maksimovic}, {Kontar}, {Zaslavsky},
  {Santolik}, {Soucek}, {Kruparova}, {Eastwood}, \& {Szabo}}]{Krupar18}
{Krupar}, V., {Maksimovic}, M., {Kontar}, E.~P., {et~al.} 2018, \apj, 857, 82

\bibitem[{{Krupar} {et~al.}(2014){Krupar}, {Maksimovic}, {Santolik}, {Kontar},
  {Cecconi}, {Hoang}, {Kruparova}, {Soucek}, {Reid}, \& {Zaslavsky}}]{Krupar14}
{Krupar}, V., {Maksimovic}, M., {Santolik}, O., {et~al.} 2014, \solphys, 289,
  3121

\bibitem[{{Krupar} {et~al.}(2020){Krupar}, {Szabo}, {Maksimovic}, {Kruparova},
  {Kontar}, {Balmaceda}, {Bonnin}, {Bale}, {Pulupa}, {Malaspina}, {Bonnell},
  {Harvey}, {Goetz}, {Dudok de Wit}, {MacDowall}, {Kasper}, {Case}, {Korreck},
  {Larson}, {Livi}, {Stevens}, {Whittlesey}, \& {Hegedus}}]{Krupar20}
{Krupar}, V., {Szabo}, A., {Maksimovic}, M., {et~al.} 2020, \apjs, 246, 57

\bibitem[{{Leblanc} {et~al.}(1998){Leblanc}, {Dulk}, \& {Bougeret}}]{Leblanc98}
{Leblanc}, Y., {Dulk}, G.~A., \& {Bougeret}, J.-L. 1998, \solphys, 183, 165

\bibitem[{{Leblanc} {et~al.}(1995){Leblanc}, {Dulk}, \& {Hoang}}]{Leblanc95}
{Leblanc}, Y., {Dulk}, G.~A., \& {Hoang}, S. 1995, \grl, 22, 3429

\bibitem[{{Maksimovic}(2019)}]{Maksimovic19}
{Maksimovic}, M. 2019, RPW-SYS-SOW-001518-LES--RPW Science Performances
  https://rpw.lesia.obspm.fr/

\bibitem[{{Maksimovic} {et~al.}(2020){Maksimovic}, {Bale}, {Chust},
  {Khotyaintsev}, {Krasnoselskikh}, {Kretzschmar}, {Plettemeier}, {Rucker},
  {Sou{\v{c}}ek}, {Steller}, {{\v{S}}tver{\'a}k}, {Tr{\'a}vn{\'\i}{\v{c}}ek},
  {Vaivads}, {Chaintreuil}, {Dekkali}, {Alexandrova}, {Astier}, {Barbary},
  {B{\'e}rard}, {Bonnin}, {Boughedada}, {Cecconi}, {Chapron}, {Chariet},
  {Collin}, {de Conchy}, {Dias}, {Gu{\'e}guen}, {Lamy}, {Leray}, {Lion},
  {Malac-Allain}, {Matteini}, {Nguyen}, {Pantellini}, {Parisot}, {Plasson},
  {Thijs}, {Vecchio}, {Fratter}, {Bellouard}, {Lorf{\`e}vre}, {Danto},
  {Julien}, {Guilhem}, {Fiachetti}, {Sanisidro}, {Laffaye}, {Gonzalez},
  {Pontet}, {Qu{\'e}ruel}, {Jannet}, {Fergeau}, {Brochot}, {Cassam-Chenai},
  {Dudok de Wit}, {Timofeeva}, {Vincent}, {Agrapart}, {Delory}, {Turin},
  {Jeandet}, {Leroy}, {Pellion}, {Bouzid}, {Katra}, {Piberne}, {Recart},
  {Santol{\'\i}k}, {Kolma{\v{s}}ov{\'a}}, {Krupa{\v{r}}},
  {Krupa{\v{r}}ov{\'a}}, {P{\'\i}{\v{s}}a}, {Uhl{\'\i}{\v{r}}}, {L{\'a}n},
  {Ba{\v{s}}e}, {Ahl{\`e}n}, {Andr{\'e}}, {Bylander}, {Cripps}, {Cully},
  {Eriksson}, {Jansson}, {Johansson}, {Karlsson}, {Puccio},
  {B{\v{r}}{\'\i}nek}, {{\"O}ttacher}, {Panchenko}, {Berthomier}, {Goetz},
  {Hellinger}, {Horbury}, {Issautier}, {Kontar}, {Krucker}, {Le Contel},
  {Louarn}, {Martinovi{\'c}}, {Owen}, {Retino}, {Rodr{\'\i}guez-Pacheco},
  {Sahraoui}, {Wimmer-Schweingruber}, {Zaslavsky}, \& {Zouganelis}}]{RPW20}
{Maksimovic}, M., {Bale}, S.~D., {Chust}, T., {et~al.} 2020, \aap, 642, A12

\bibitem[{{Maksimovic} \& {et al.}(2021)}]{Maksimovic1st}
{Maksimovic}, M. \& {et al.} 2021, \aap

\bibitem[{{Manning} \& {Fainberg}(1980)}]{Manning80}
{Manning}, R. \& {Fainberg}, J. 1980, Space Science Instrumentation, 5, 161

\bibitem[{{Meyer-Vernet} {et~al.}(2017){Meyer-Vernet}, {Issautier}, \&
  {Moncuquet}}]{Meyer17}
{Meyer-Vernet}, N., {Issautier}, K., \& {Moncuquet}, M. 2017, Journal of
  Geophysical Research (Space Physics), 122, 7925

\bibitem[{{M{\"u}ller} {et~al.}(2020){M{\"u}ller}, {St. Cyr}, {Zouganelis},
  {Gilbert}, {Marsden}, {Nieves-Chinchilla}, {Antonucci}, {Auch{\`e}re},
  {Berghmans}, {Horbury}, {Howard}, {Krucker}, {Maksimovic}, {Owen}, {Rochus},
  {Rodriguez-Pacheco}, {Romoli}, {Solanki}, {Bruno}, {Carlsson}, {Fludra},
  {Harra}, {Hassler}, {Livi}, {Louarn}, {Peter}, {Sch{\"u}hle}, {Teriaca}, {del
  Toro Iniesta}, {Wimmer-Schweingruber}, {Marsch}, {Velli}, {De Groof},
  {Walsh}, \& {Williams}}]{Muller2020}
{M{\"u}ller}, D., {St. Cyr}, O.~C., {Zouganelis}, I., {et~al.} 2020, \aap, 642,
  A1

\bibitem[{{Musset} {et~al.}(2021){Musset}, {Maksimovic}, {Kontar}, {Krupar},
  {Chrysaphi}, {Bonnin}, {Vecchio}, {Cecconi}, {Bale}, \& {Pulupa}}]{Musset21}
{Musset}, S., {Maksimovic}, M., {Kontar}, E., {et~al.} 2021, \aap, Submitted

\bibitem[{{Novaco} \& {Brown}(1978)}]{Novaco_Brown78}
{Novaco}, J.~C. \& {Brown}, L.~W. 1978, \apj, 221, 114

\bibitem[{{Panchenko}(2017)}]{Panchenko17}
{Panchenko}, M. 2017, Scientific Report on FFG/ASAP 11 project SOLOCAL,
  Projektnummer: 847978

\bibitem[{{Rodr{\'\i}guez-Pacheco} {et~al.}(2020){Rodr{\'\i}guez-Pacheco},
  {Wimmer-Schweingruber}, {Mason}, {Ho}, {S{\'a}nchez-Prieto}, {Prieto},
  {Mart{\'\i}n}, {Seifert}, {Andrews}, {Kulkarni}, {Panitzsch}, {Boden},
  {B{\"o}ttcher}, {Cernuda}, {Elftmann}, {Espinosa Lara}, {G{\'o}mez-Herrero},
  {Terasa}, {Almena}, {Begley}, {B{\"o}hm}, {Blanco}, {Boogaerts}, {Carrasco},
  {Castillo}, {da Silva Fari{\~n}a}, {de Manuel Gonz{\'a}lez}, {Drews},
  {Dupont}, {Eldrum}, {Gordillo}, {Guti{\'e}rrez}, {Haggerty}, {Hayes},
  {Heber}, {Hill}, {J{\"u}ngling}, {Kerem}, {Knierim}, {K{\"o}hler}, {Kolbe},
  {Kulemzin}, {Lario}, {Lees}, {Liang}, {Mart{\'\i}nez Hell{\'\i}n}, {Meziat},
  {Montalvo}, {Nelson}, {Parra}, {Paspirgilis}, {Ravanbakhsh}, {Richards},
  {Rodr{\'\i}guez-Polo}, {Russu}, {S{\'a}nchez}, {Schlemm}, {Schuster},
  {Seimetz}, {Steinhagen}, {Tammen}, {Tyagi}, {Varela}, {Yedla}, {Yu},
  {Agueda}, {Aran}, {Horbury}, {Klecker}, {Klein}, {Kontar}, {Krucker},
  {Maksimovic}, {Malandraki}, {Owen}, {Pacheco}, {Sanahuja}, {Vainio},
  {Connell}, {Dalla}, {Dr{\"o}ge}, {Gevin}, {Gopalswamy}, {Kartavykh},
  {Kudela}, {Limousin}, {Makela}, {Mann}, {{\"O}nel}, {Posner}, {Ryan},
  {Soucek}, {Hofmeister}, {Vilmer}, {Walsh}, {Wang}, {Wiedenbeck}, {Wirth}, \&
  {Zong}}]{Pacheco20}
{Rodr{\'\i}guez-Pacheco}, J., {Wimmer-Schweingruber}, R.~F., {Mason}, G.~M.,
  {et~al.} 2020, \aap, 642, A7

\bibitem[{Zarka {et~al.}(2004)Zarka, Cecconi, \& Kurth}]{Zarka04}
Zarka, P., Cecconi, B., \& Kurth, W.~S. 2004, Journal of Geophysical Research:
  Space Physics, 109

\bibitem[{{Zaslavsky} {et~al.}(2011){Zaslavsky}, {Meyer-Vernet}, {Hoang},
  {Maksimovic}, \& {Bale}}]{Zaslavsky11}
{Zaslavsky}, A., {Meyer-Vernet}, N., {Hoang}, S., {Maksimovic}, M., \& {Bale},
  S.~D. 2011, Radio Science, 46, RS2008

\bibitem[{{Zouganelis} {et~al.}(2020){Zouganelis}, {De Groof}, {Walsh},
  {Williams}, {M{\"u}ller}, {St Cyr}, {Auch{\`e}re}, {Berghmans}, {Fludra},
  {Horbury}, {Howard}, {Krucker}, {Maksimovic}, {Owen},
  {Rodr{\'\i}guez-Pacheco}, {Romoli}, {Solanki}, {Watson}, {Sanchez}, {Lefort},
  {Osuna}, {Gilbert}, {Nieves-Chinchilla}, {Abbo}, {Alexandrova},
  {Anastasiadis}, {Andretta}, {Antonucci}, {Appourchaux}, {Aran}, {Arge},
  {Aulanier}, {Baker}, {Bale}, {Battaglia}, {Bellot Rubio}, {Bemporad},
  {Berthomier}, {Bocchialini}, {Bonnin}, {Brun}, {Bruno}, {Buchlin},
  {B{\"u}chner}, {Bucik}, {Carcaboso}, {Carr}, {Carrasco-Bl{\'a}zquez},
  {Cecconi}, {Cernuda Cangas}, {Chen}, {Chitta}, {Chust}, {Dalmasse},
  {D'Amicis}, {Da Deppo}, {De Marco}, {Dolei}, {Dolla}, {Dudok de Wit}, {van
  Driel-Gesztelyi}, {Eastwood}, {Espinosa Lara}, {Etesi}, {Fedorov},
  {F{\'e}lix-Redondo}, {Fineschi}, {Fleck}, {Fontaine}, {Fox}, {Gandorfer},
  {G{\'e}not}, {Georgoulis}, {Gissot}, {Giunta}, {Gizon}, {G{\'o}mez-Herrero},
  {Gontikakis}, {Graham}, {Green}, {Grundy}, {Haberreiter}, {Harra}, {Hassler},
  {Hirzberger}, {Ho}, {Hurford}, {Innes}, {Issautier}, {James}, {Janitzek},
  {Janvier}, {Jeffrey}, {Jenkins}, {Khotyaintsev}, {Klein}, {Kontar},
  {Kontogiannis}, {Krafft}, {Krasnoselskikh}, {Kretzschmar}, {Labrosse},
  {Lagg}, {Landini}, {Lavraud}, {Leon}, {Lepri}, {Lewis}, {Liewer}, {Linker},
  {Livi}, {Long}, {Louarn}, {Malandraki}, {Maloney}, {Martinez-Pillet},
  {Martinovic}, {Masson}, {Matthews}, {Matteini}, {Meyer-Vernet}, {Moraitis},
  {Morton}, {Musset}, {Nicolaou}, {Nindos}, {O'Brien}, {Orozco Suarez},
  {Owens}, {Pancrazzi}, {Papaioannou}, {Parenti}, {Pariat}, {Patsourakos},
  {Perrone}, {Peter}, {Pinto}, {Plainaki}, {Plettemeier}, {Plunkett}, {Raines},
  {Raouafi}, {Reid}, {Retino}, {Rezeau}, {Rochus}, {Rodriguez},
  {Rodriguez-Garcia}, {Roth}, {Rouillard}, {Sahraoui}, {Sasso}, {Schou},
  {Sch{\"u}hle}, {Sorriso-Valvo}, {Soucek}, {Spadaro}, {Stangalini}, {Stansby},
  {Steller}, {Strugarek}, {{\v{S}}tver{\'a}k}, {Susino}, {Telloni}, {Terasa},
  {Teriaca}, {Toledo-Redondo}, {del Toro Iniesta}, {Tsiropoula}, {Tsounis},
  {Tziotziou}, {Valentini}, {Vaivads}, {Vecchio}, {Velli}, {Verbeeck},
  {Verdini}, {Verscharen}, {Vilmer}, {Vourlidas}, {Wicks},
  {Wimmer-Schweingruber}, {Wiegelmann}, {Young}, \& {Zhukov}}]{Zouganelis_SAP}
{Zouganelis}, I., {De Groof}, A., {Walsh}, A.~P., {et~al.} 2020, \aap, 642, A3

\end{thebibliography}

\newpage 

\appendix
\section{Calibration using the Galaxy}
 \label{gal_cal}
 The presence of a strong time-varying EM contamination at 120 kHz and 80 kHz and its harmonics that is due to the PCDU radiated by the Solar Panels and the reaction wheel electronic box, respectively, makes the observation of the galactic background much more difficult as it is masked by the instrumental noise \citep{Maksimovic1st}. {However, to reduce the effect of the EM contamination and infer the Galaxy signal from the data, we performed an analysis over about ten months of HFR measurements. This analysis allowed us to estimate the reference system noise of HFR in-flight when RPW is embedded in the space plasma environment, and to derive a value for $\Gamma l_{eff} $ for the V1-V2 dipole. The latter is compatible, within the uncertainty limit, with the  $\Gamma l_{eff} $ value obtained from the cross-calibration with the type III emission measured by Wind. }
\subsection{System noise in space}
As with previous missions, such as Stereo \citep{Zaslavsky11}, it is necessary to determine the instrument system noise in space that has to be removed in order to retrieve, for instance, the physical ambient noise due to the Galaxy.  
{Unfortunately, there were no measurements available with all three antennas undeployed due to a spacecraft shutdown and loss of scientific data. About $1.5$ hours of observation with TNR-HFR were carried out two days after the launch at the instrument second switch-on (12 February 2020, 13:26 to 16:00 UT), following the deployment of the antenna Pz (sensor V1 of TNR-HFR). }
A series of 164 and 329 spectra, covering the full spectral range of the receiver, were recorded by TNR and HFR  connected to all the available sensors. 
Figure \ref{backg12} shows the TNR-HFR backgrounds for the dipole sensors measured before the deployment of antennas V2 and V3. These backgrounds have been obtained by considering the minimum level detected at each frequency during this 1.5-hour interval. As expected, since V1 is from a deployed antenna, the pre-deployment  background levels for sensors V1-V2 and V3-V1 are higher with respect to V2-V3. The high frequency part ($>$ 5MHz) of the HFR V2-V3  shows an increase in power with the presence of large peaks at 8 and 15 MHz, which is probably due to some coupling with the spacecraft.
Moreover, all HFR  backgrounds show several interference line due to the spacecraft and TNR backgrounds show interference lines only in the high frequency part of the spectrum. Since we are only interested in the receiver's noise level (to be used as a reference), the interference lines should be removed. For this purpose, we identified the lower envelopes of of TNR and HFR V2-V3 backgrounds via a linear interpolation among the absolute minima, thus removing discrete interference lines (with linear interpolation through them). The resulting background noise is shown in Figure (\ref{backg12}) (thick dashed lines).
\begin{figure*}
\centering{
\includegraphics[scale=0.5]{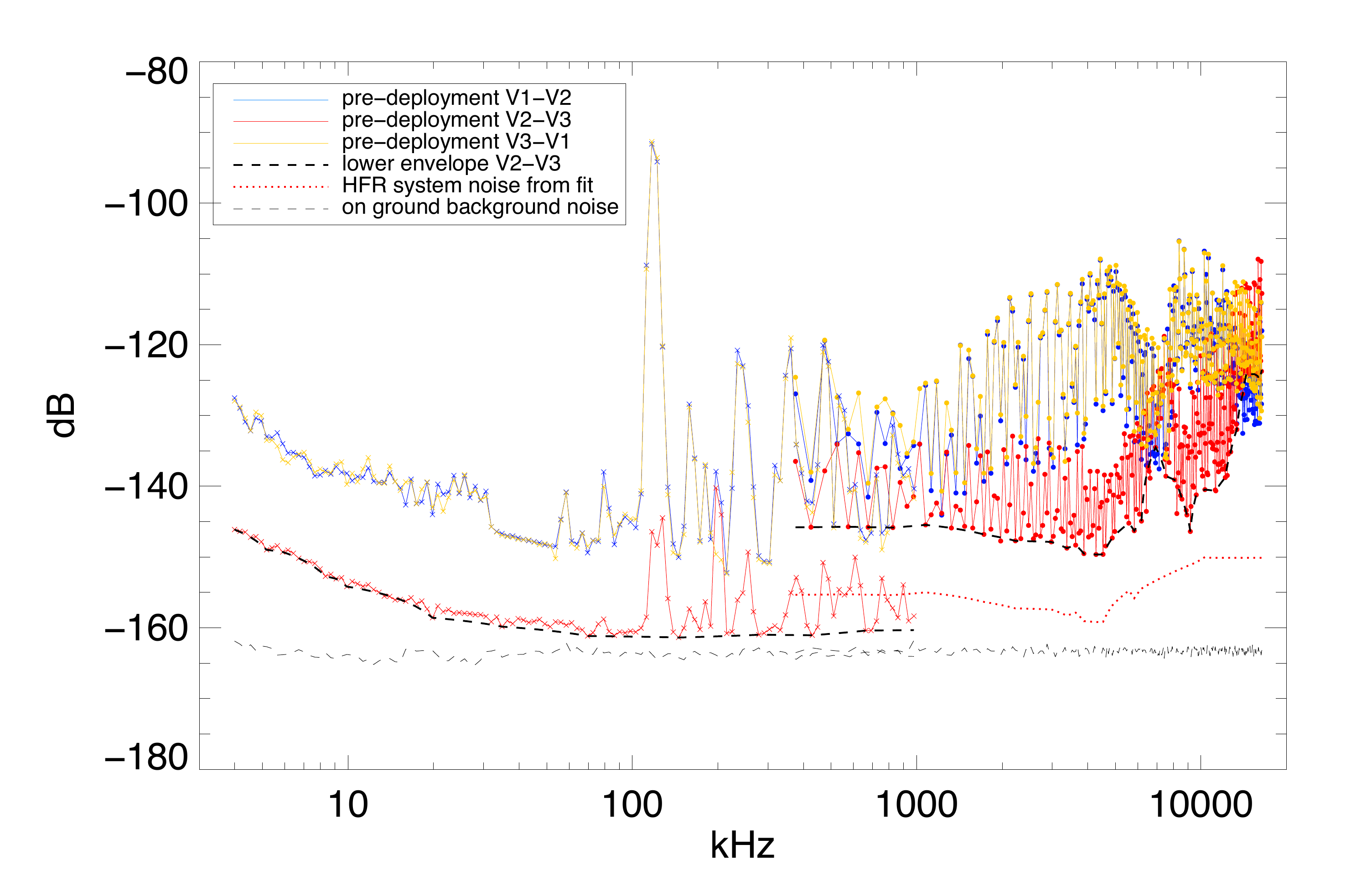}
}
\caption{Measurements of the pre-deployment voltage spectral density (in dB) of TNR (crosses) and HFR (dots) to evaluate the receivers system noise for the three dipole sensors {V1-V2 (blue), V2-V3 (red) and V3-V1 (orange)}. 
Thick black dashed lines correspond to the lower envelope of the background spectral density representing the reference receiver pre-deployment background. Thin black dashed lines correspond to the background noise measured during the on-ground tests in Meudon and including the effects of the preamplifiers. The red dotted line indicates the HFR system noise obtained after the fitting procedure described in Sect. \ref{Sect_gal_back}.
 }
\label{backg12}
\end{figure*}
\subsection{Galactic background}
\label{Sect_gal_back}
The presence of the  EM contamination makes the observation of the galactic background very difficult. Indeed,  the HFR spectrum measured after the deployment of the RPW antennas,  appears sawtoothed with the power of peaks and troughs varying in time. 
To reduce the effects of the interference, we analyzed HFR  data over a long time period of about ten months.
\begin{figure}
\centering{
\includegraphics[scale=0.3]{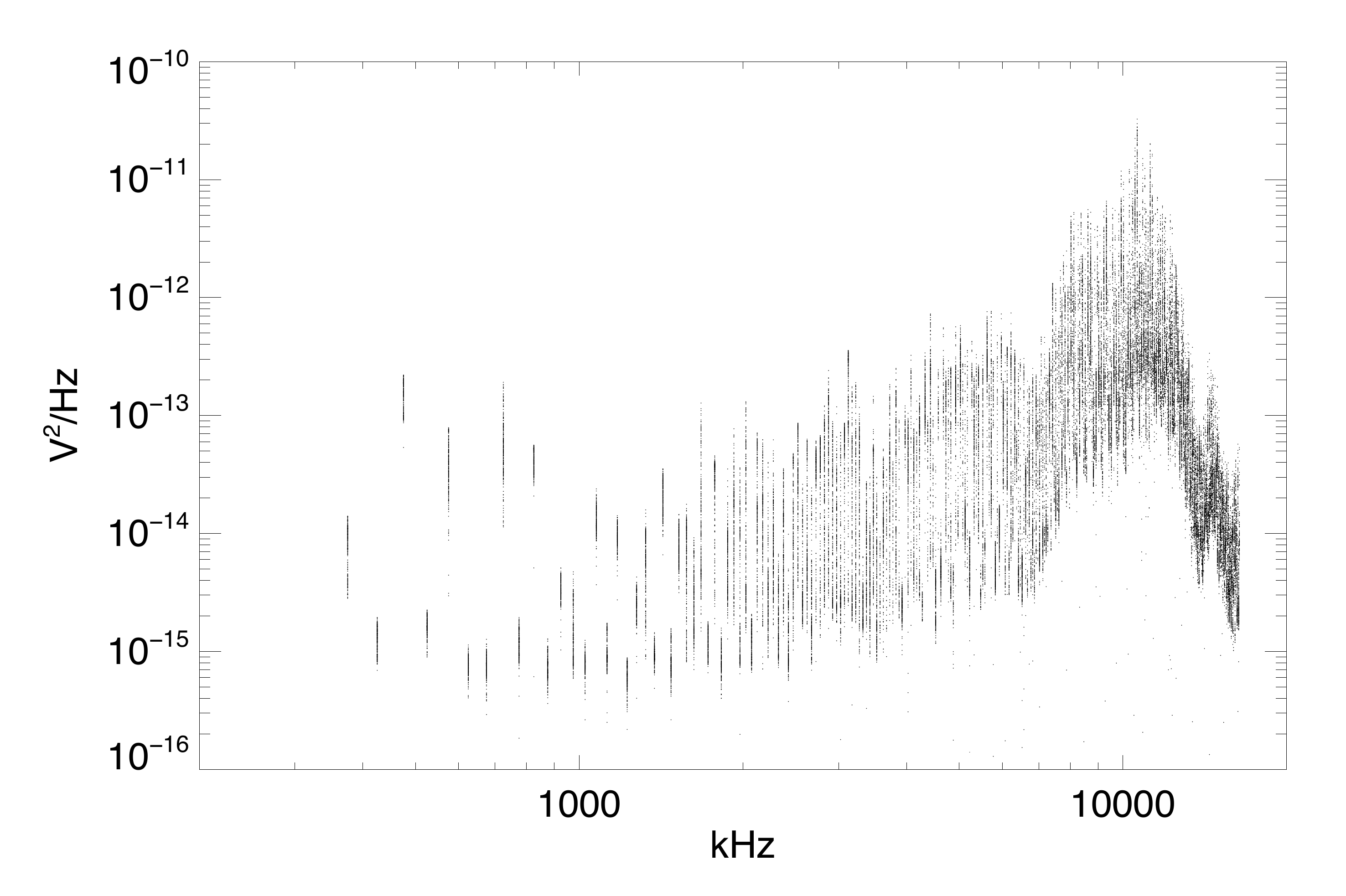}
}
\caption{{Daily minima of the HFR spectra over the time span from March 1 to December 31, 2020 for the sensor V1-V2 as a function of the frequency.}
}
\label{HFR_full}
\end{figure}
The observed spectrum, which corresponds to the sum of galactic background noise and receiver noise, is derived as follows. We first considered all the V1-V2 spectra acquired in the period between March 1 and December 31, 2020 (unlike the other dipoles, V1-V2 are available for every day) and we then evaluate the daily minima for each frequency (Figure \ref{HFR_full}).  A forest of interference lines due to the harmonics of 80 and 120 kHz is still visible in the data.  At each frequency, the statistical fluctuations of the background level appear as a well-defined gaussian distribution \citep{Zarka04} centered on the average value of $V^{2}_{gal}+V^{2}_{system-noise}$ at that frequency (an example at a given frequency is shown in Figure \ref{histo_hfr}).
\begin{figure}
\centering{
\includegraphics[scale=0.3]{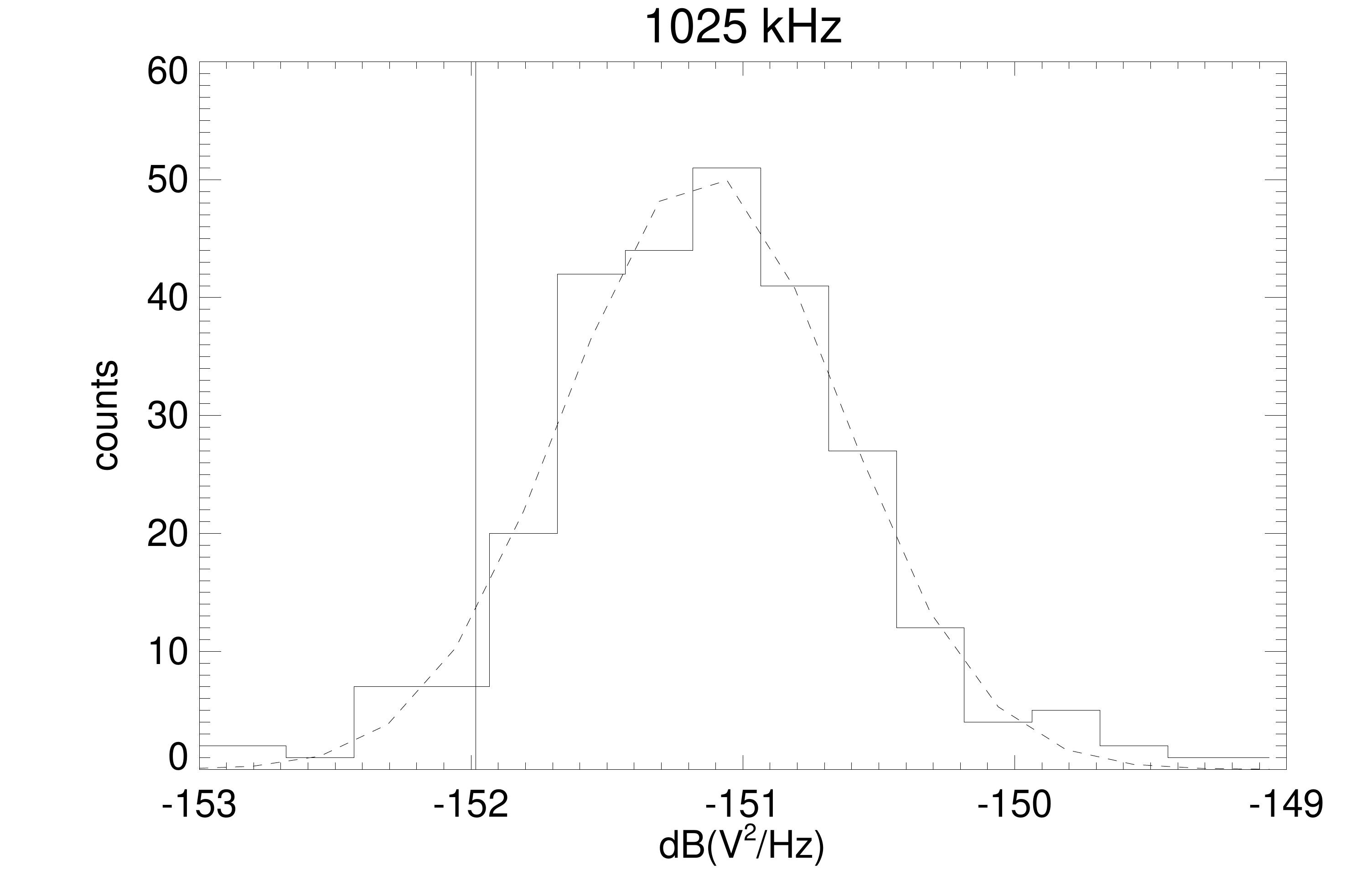}
}
\caption{Histogram of the minima values in the period between March {1} and December 31, 2020 at the frequency of 1.025 MHz.  The gaussian distribution (dashed) corresponds to the Galactic signal plus the system noise and their fluctuations at this frequency. The vertical line indicates the limit of the  lower $5\%$ occurrence level.}
\label{histo_hfr}
\end{figure}
Similarly to what was considered for the calibration of other receivers in space \citep{Zarka04} we considered the lower $5\%$ occurrence level detected at each frequency as our reference for $V^{2}_{gal}+V^{2}_{system-noise}$ (black line in figure \ref{gal_noise}).
\begin{figure}
\centering{
\includegraphics[scale=0.3]{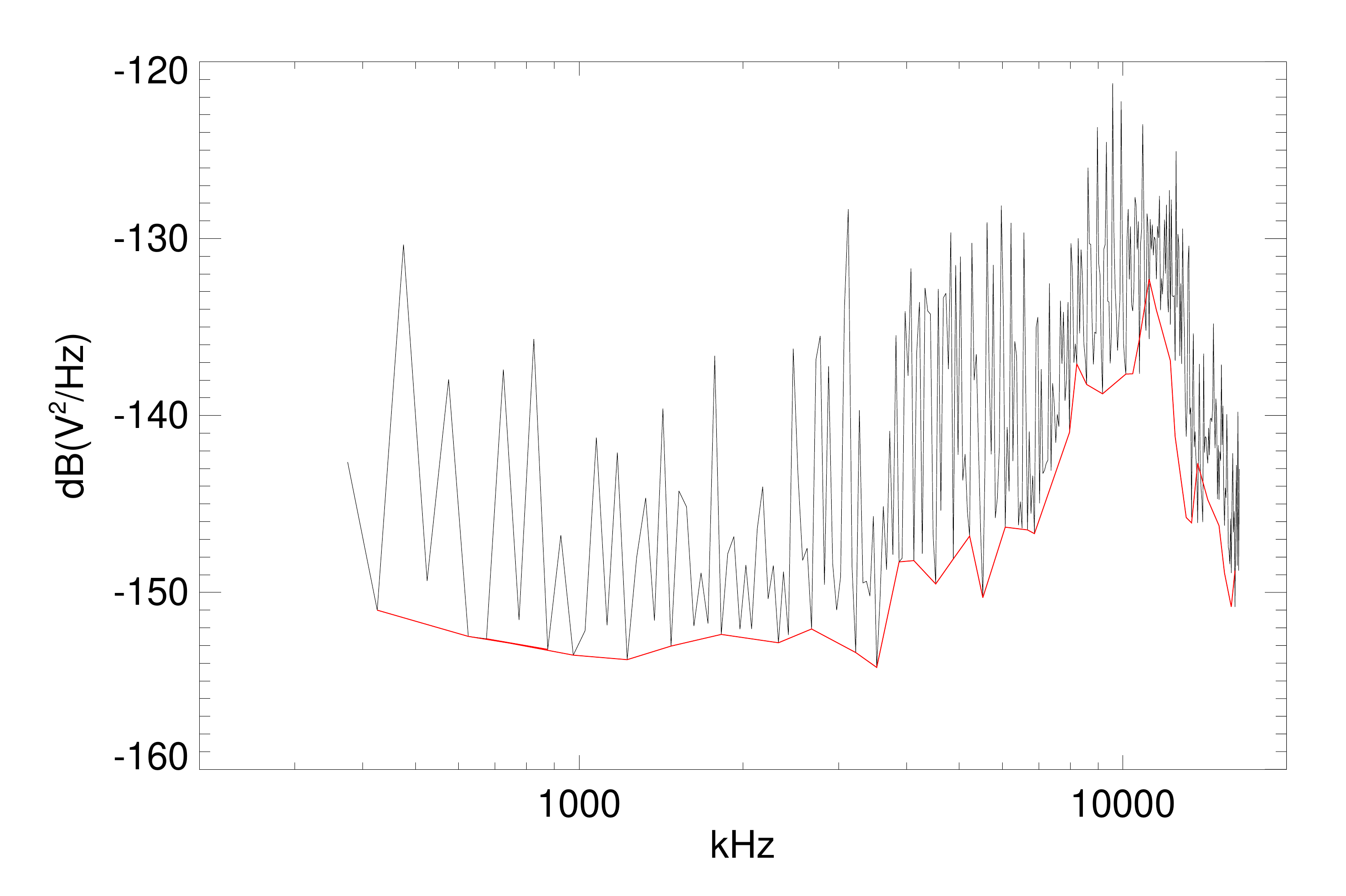}
}
\caption{ {Lower $5\%$ occurrence level detected at each frequency of Figure \ref{HFR_full} (black line). Lower envelope representing the reference for $V^{2}_{gal}+V^{2}_{noise}$ (red line).} }
\label{gal_noise}
\end{figure}
As we do for defining the system noise, we remove the effect of the spacecraft interference lines from the galactic background by considering only the  lower envelope (red line in figure \ref{gal_noise}).

The found minimum background can be compared to a sky galactic radio background model. In this work, we use the model from \citet{Novaco_Brown78}, which is an empirical model for the isotropic sky background brightness based on space measurements. The model is expressed as:
\begin{equation}
B_{model}= B_0f_{MHz}^{-0.76}e^{-\tau}
\label{eq_novaco}
,\end{equation}
where $f_{MHz}$  is the frequency in MHz, $\tau = 3.28f_{MHz}^{-0.64}$ and $B = 1.38 \times 10^{-19}$ W/m$^2$/Hz/sr. 
By reversing Equation (\ref{eq1gal}), the product $\Gamma l_{eff}$ can be obtained as:
\begin{equation}
\Gamma l_{eff}=\left(\frac{3}{4\pi Z_0}\frac{V^2_r-V^2_{noise}-\Gamma^2V^2_{QTN}}{B_{model}} \right)^{1/2} \;\; .
\label{gammagal}
\end{equation}
The value of $ \Gamma l_{eff}$ is determined by taking the average of the right-hand side of Equation (\ref{gammagal}) in the frequency range where it is almost constant. The analysis is thus restricted to the frequency range 1.2-3.1 MHz, chosen in such a way that  the short dipole approximation is valid ($f<$ 4 MHz) and the plasma thermal noise, the term $\Gamma^2V^2_{QTN}$, can be neglected ($f>$ 1.2 MHz). \\
Since the level of the galactic background signal is very low and close to the receiver noise, the determination of $V_{noise}^2$ is crucial for a very accurate antenna calibration. 
As was already the case for previous instrumentations of this kind \citep{Zaslavsky11}, it is not possible to directly measure  the precise system noise level for HFR. We thus assume that the actual system noise  level is between the pre-deployment HFR spectrum (red line in Figure (\ref{backg12}) ) and the HFR background measured on the ground, including the preamplifiers (thin dashed line).
{Thus, to get an accurate value of $\Gamma l_{eff}$, $V^2_{noise}$ has been varied between the two curves by constant steps. Operationally this is done by iteratively reducing the system noise, starting at the measured pre-deployment HFR spectrum, by 0.2 dB until the lowest level, given by the on-ground background, is reached.} For each value of $V^2_{noise}$ we evaluate the mean value of $\Gamma l_{eff}$ between 1.2-3.1 MHz by using Equation (\ref{gammagal}) and the $\chi^2$ of the data points with respect to this mean \citep{Zaslavsky11}.  
We then retain the values of $\Gamma l_{eff}$ that minimize the $\chi^2$} , the uncertainty on $\Gamma l_{eff}$ being defined as $\sqrt{\chi^2/N_{data}}$. The best fit for the sensor V1-V2 is shown in Figure \ref{Gleff_fit}. The fitting procedure provides $\Gamma l_{eff}=3.4\pm0.1$ and an average value of the noise, in the range of frequency 1.2-3.1 MHz, of -156.9 dB(V$^2$/Hz) that is 2$\cdot$10$^{-16}$ V$^2$/Hz. 
\begin{figure}
\centering{
\includegraphics[scale=0.3]{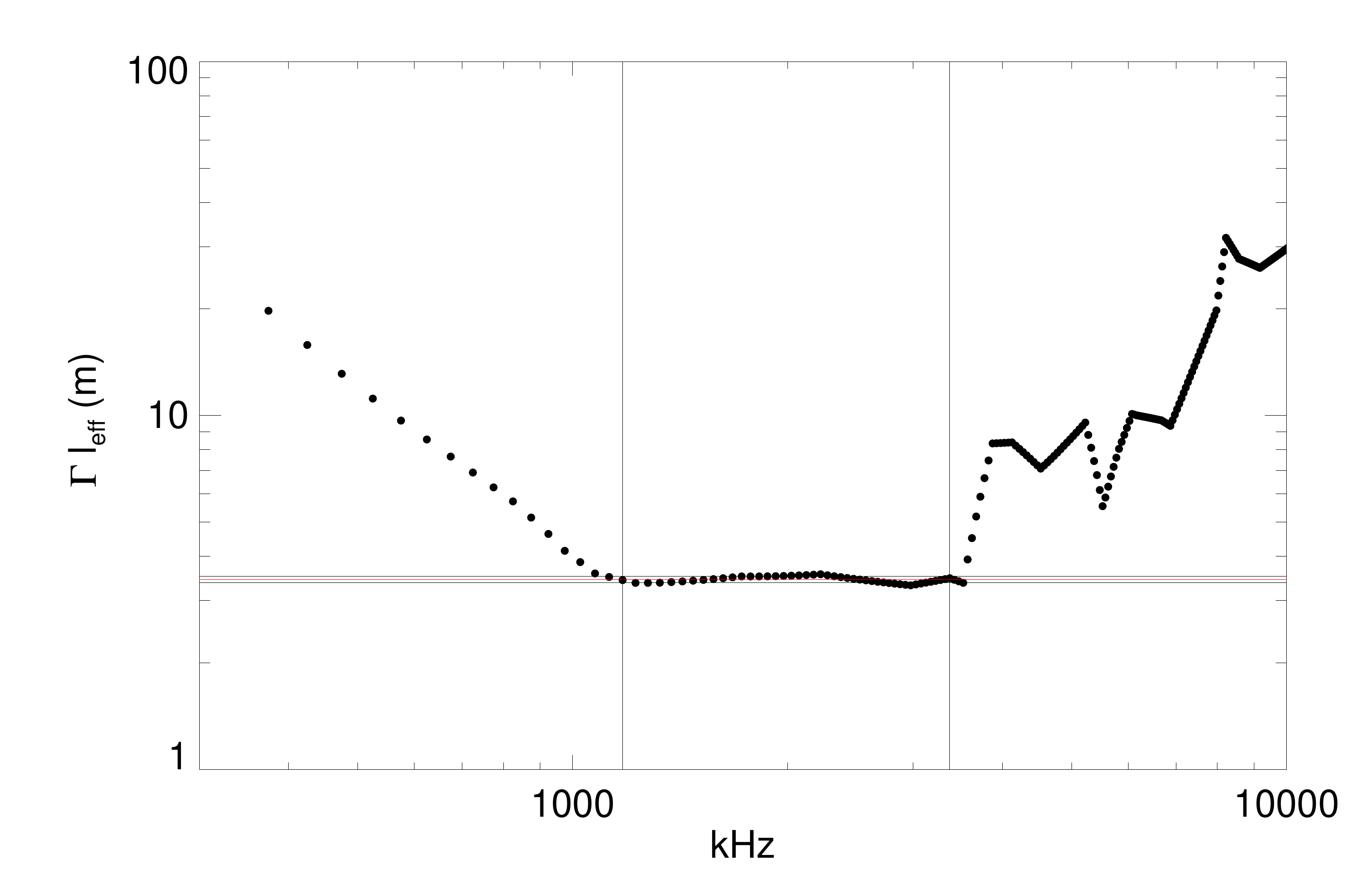}
}
\caption{Depiction of the right-hand side of Equation (\ref{gammagal}) as a function of the frequency. The horizontal lines show the mean
value (red) and $1\sigma$ uncertainties (black) calculated in the frequency range between the two dashed vertical lines for
the minimum $\chi^2$ fit.}
\label{Gleff_fit}
\end{figure}
{Since  for $f<$1.2 MHz and $f>$3.1 MHz the receiver noise cannot be measured through the Galaxy signal, we assume that the value of the total downward shift $\Delta B\sim$ 14 dB, allowing to get the best fit in the 1.2-3.1 MHz range, also applies to the full HFR band. The $V^{2}_{system-noise}$  for HFR, representing the true system noise level measured in space where the antennas and pre-amplifiers are embedded in the solar wind plasma, is  obtained by subtracting $\Delta B$ to the measured pre-deployment HFR spectrum and it is shown in figure \ref{backg12} (red dotted line)}.  We remark  that at frequencies above 5 MHz,  the  peaks due to the possible coupling with the spacecraft have been removed by means of linear interpolations.
The value of $\Gamma l_{eff}$ for the dipole antenna $V1-V2,$ as obtained from the Galactic background signal, is shown in in Table \ref{tab0}. We note that the values found by using  independent methods -- that is, the comparison with the type III radio power as measured by Wind/Waves on March 30 and the galactic background {inferred by using ten months of HFR measurements} -- are compatible within the uncertainty limits. 
Figure \ref{gal_power} (upper panel) shows the comparison of the signal $V^2_r-V^2_{system-noise}$, calculated from the minimum background, with the theoretical  prediction based on the \citep{Novaco_Brown78} model and Equation (\ref{eq1gal}) in which $l_{eff}$ is  given by the minimum $\chi^2$ fit. Figure \ref{gal_power} shows a good agreement between the data and the model in the short dipole frequency range: between 1.2-3.1 MHz, the relative error between the data and the model curve is less than 5\%. 
\begin{figure}
\centering{
\vbox{
\includegraphics[scale=0.3]{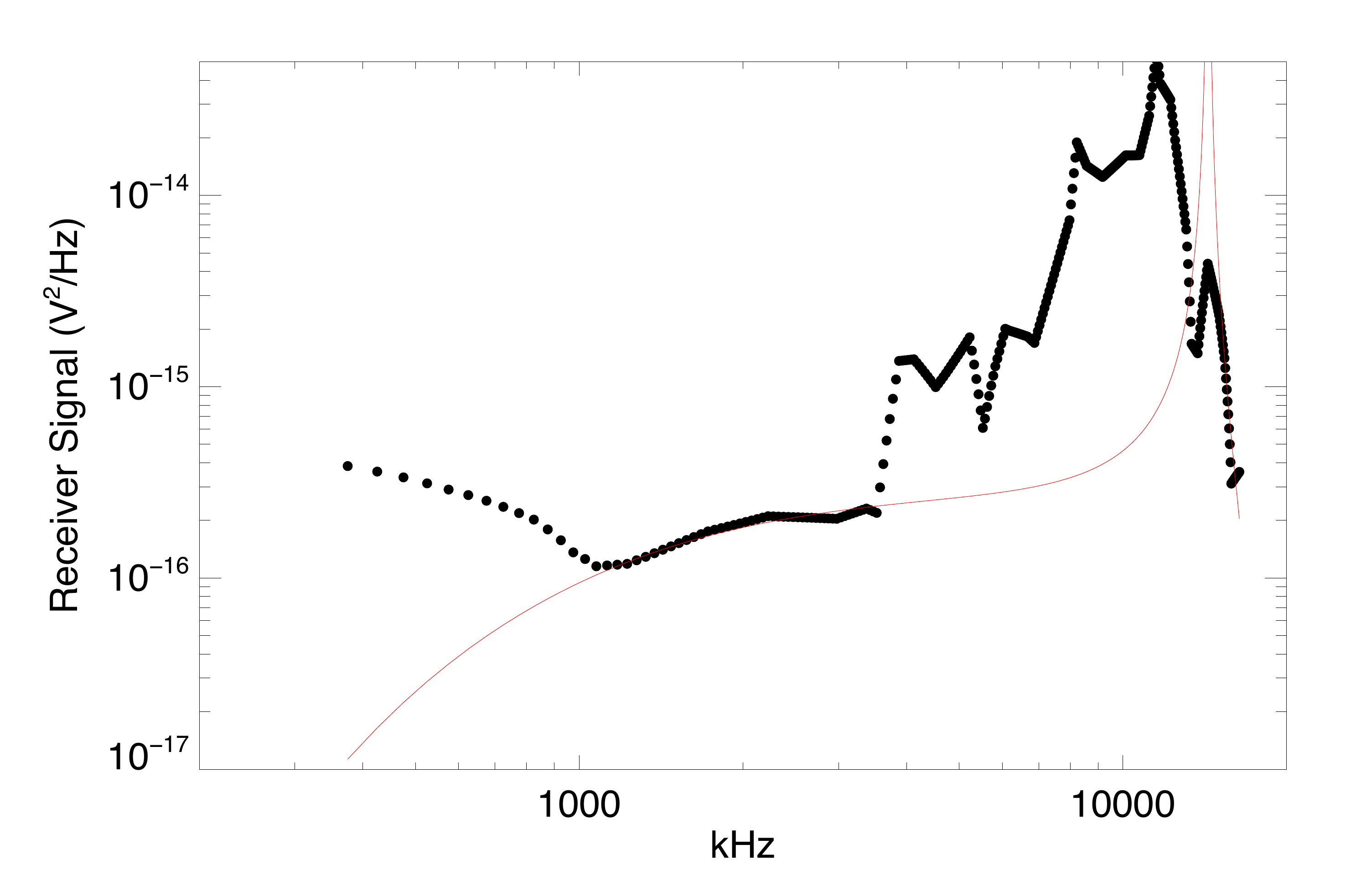}
\includegraphics[scale=0.3]{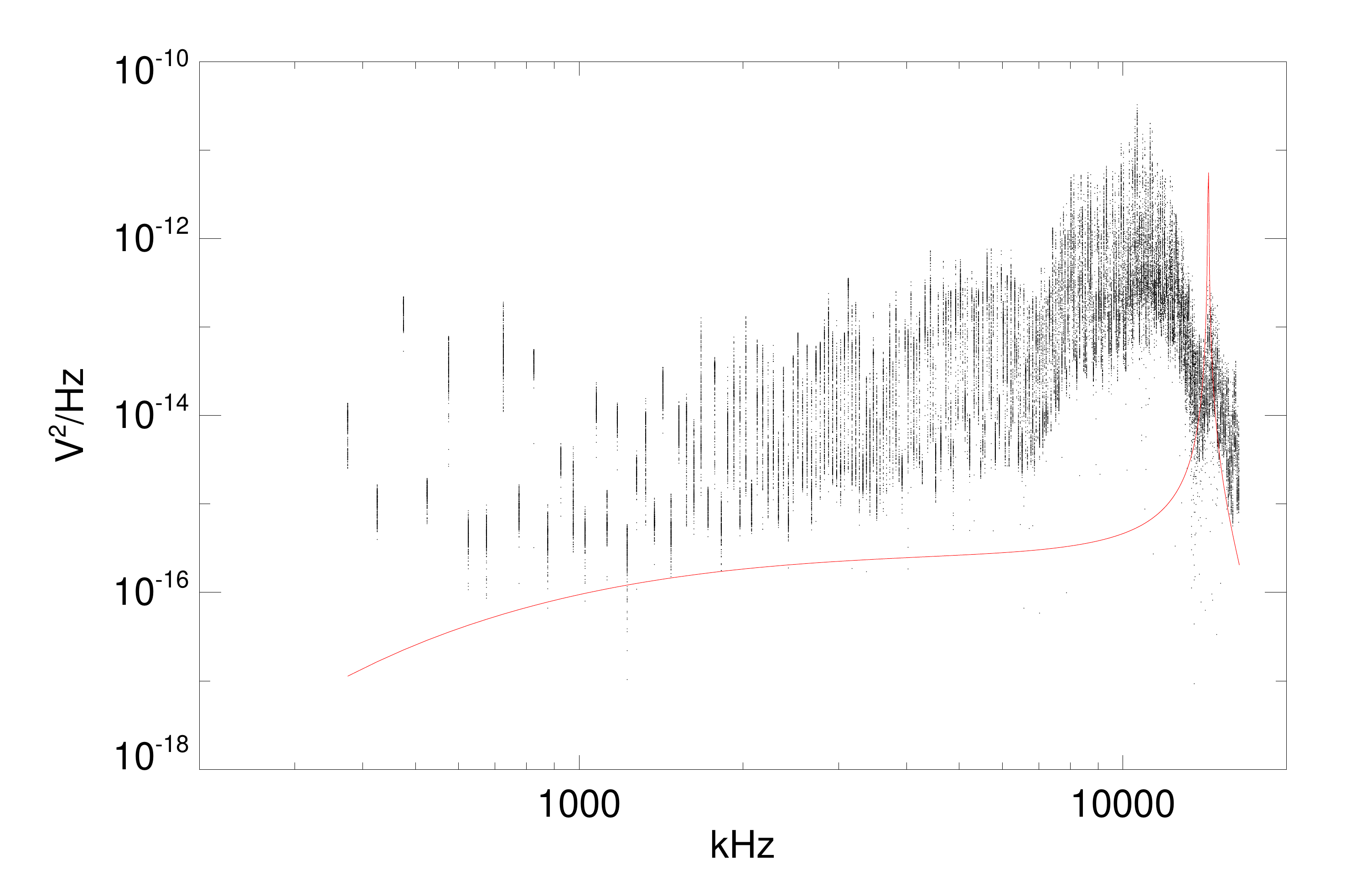}
}
}
\caption{{Comparison of the measured receiver power spectral density with the theoretical received power. Upper: $V^2_r-V^2_{system-noise}$ for the sensor V1-V2, as a function of the frequency, calculated from the minimum background. Black dots represent the data points. Lower: daily minima of the HFR $V^2_r$ spectra minus $V^2_{system-noise}$ as a function of the frequency, over the time span March 1-December 31, 2020 for the sensor V1-V2.} Red solid line represents the theoretical receiver power, expected for the model \citet{Novaco_Brown78}, as obtained from Equations (\ref{eq1gal}), (\ref{eq_novaco}),(\ref{gammagal}) using $\Gamma l_{eff}=3.4\pm0.1$ m (from the galactic background measurements). The increase of the measured voltage spectral power at frequencies above 4 MHz is the consequence of the antenna resonance discussed in the text.}
\label{gal_power}
\end{figure}

Figure \ref{gal_power} (lower panel) shows the comparison among daily minima of the receiver power spectral density $V^2_r-V^2_{system-noise}$  over the time span March 1-December 31, 2020 for the sensor $V1-V2$ as a function of the frequency and the theoretical receiver power from Equations (\ref{eq1gal}), (\ref{eq_novaco}),and (\ref{gammagal}), calculated by using $l_{eff}=7.7\pm0.3$ m, as obtained from the galactic background and derived in Section \ref{antenne}.
\begin{figure}
\centering{
\includegraphics[scale=0.3]{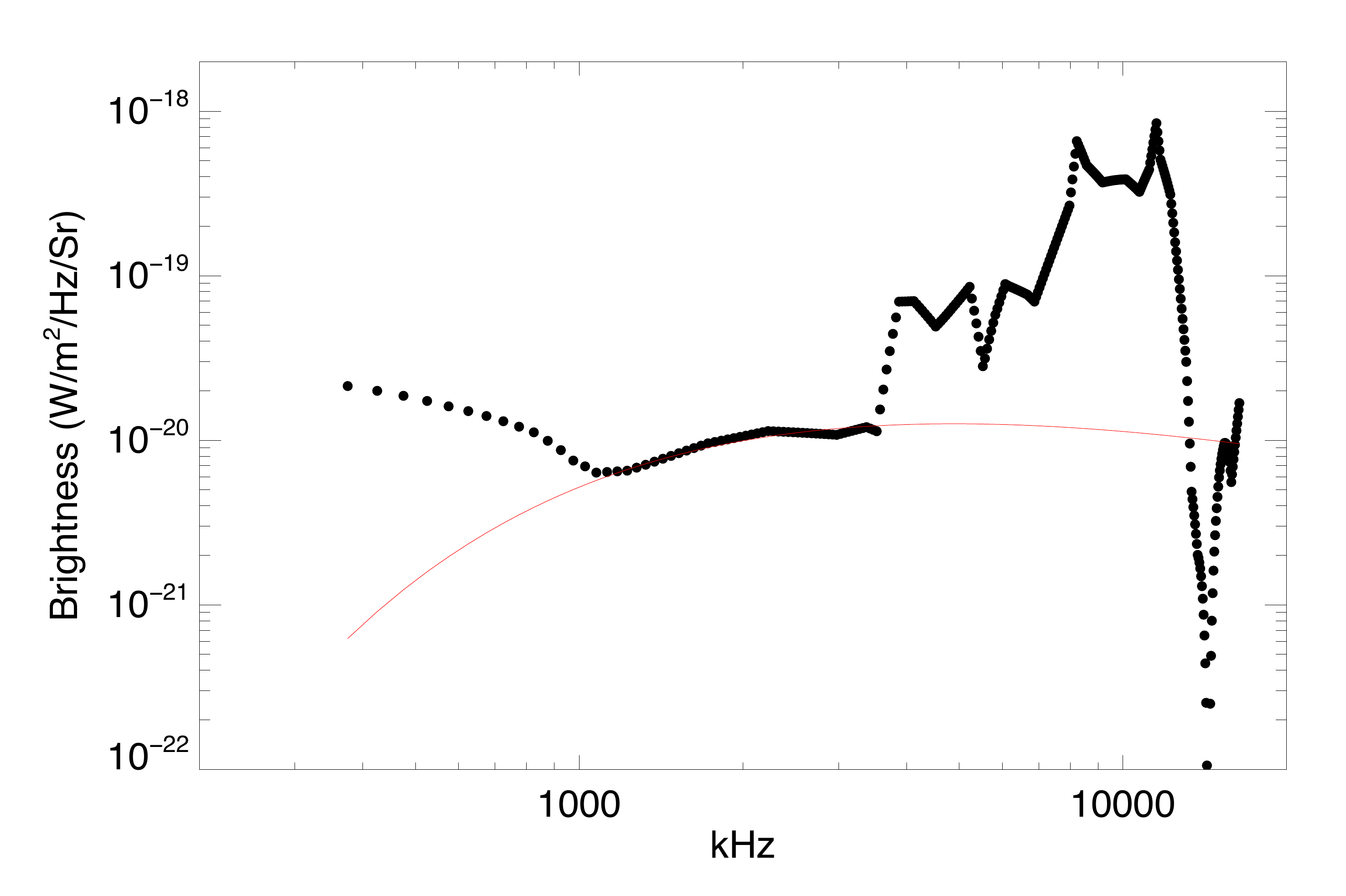}
}
\caption{Calibrated data expressed in brightness as a function of the frequency (black dots). The galactic background radiation model from \citet{Novaco_Brown78} (red solid line) is overplotted.}
\label{gal_brig}
\end{figure}
 For the purposes of illustration and comparison, Figure (\ref{gal_brig}) shows the model of galactic background together with the receiver background data as functions of frequency, calibrated in W/m$^2$/Hz/Sr.

\end{document}